\documentclass[twocolumn]{aastex61}
\usepackage{mathrsfs, mathtools}
\definecolor{darkbrown}{RGB}{100,50,30}

\begin{document}

\title{Homogeneous Analysis of the Dust Morphology of Transition Disks Observed with ALMA: Investigating dust trapping and the origin of the cavities}

\correspondingauthor{Paola~Pinilla, Hubble Fellow}
\affiliation{Department of Astronomy/Steward Observatory, The University of Arizona, 933 North Cherry Avenue, Tucson, AZ 85721, USA}
\email{pinilla@email.arizona.edu}

\author{P.~Pinilla}
\affiliation{Department of Astronomy/Steward Observatory, The University of Arizona, 933 North Cherry Avenue, Tucson, AZ 85721, USA}

\author{M.~Tazzari}
\affiliation{Institute of Astronomy, University of Cambridge, Madingley Road, Cambridge CB3 0HA, UK}

\author{I.~Pascucci}
\affiliation{Lunar and Planetary Laboratory, The University of Arizona, Tucson, AZ 85721, USA}
\affiliation{Earths in Other Solar Systems Team, NASA Nexus for Exoplanet System Science}

\author{A.~N.~Youdin}
\affiliation{Department of Astronomy/Steward Observatory, The University of Arizona, 933 North Cherry Avenue, Tucson, AZ 85721, USA}

\author{A.~Garufi}
\affiliation{Universidad Aut\'{o}noma de Madrid, Dpto. F\'{i}sica Te\'{o}rica, M\'{o}dulo 15, Facultad de Ciencias, Campus de Cantoblanco, E-28049 Madrid, Spain}

\author{C.~F.~Manara}
\affiliation{European Southern Observatory, Karl-Schwarzschild-Str. 2, D-85748 Garching, Germany}

\author{L.~Testi}
\affiliation{European Southern Observatory, Karl-Schwarzschild-Str. 2, D-85748 Garching, Germany}
\affiliation{INAF-Arcetri, Largo E. Fermi 5, I-50125 Firenze, Italy}

\author{G.~van~der~Plas}
\affiliation{Univ. Grenoble Alpes, CNRS, IPAG, F-38000 Grenoble, France}

\author{S.~A.~Barenfeld}
\affiliation{California Institute of Technology, Department of Astronomy, MC 249-17, Pasadena, CA 91125, USA}

\author{H.~Canovas}
\affiliation{European Space Astronomy Centre (ESA), P.O. Box, 78, E-28691 Villanueva de la Ca\~{n}ada, Madrid, Spain}

\author{E.~G.~Cox}
\affiliation{Department of Astronomy, University of Illinois at Urbana-Champaign, 1002 W. Green Street, Urbana, IL 61801, USA}

\author{N.~P.~Hendler}
\affiliation{Lunar and Planetary Laboratory, The University of Arizona, Tucson, AZ 85721, USA}
\affiliation{LSSTC Data Science Fellow}

\author{L.~M.~P\'erez}
\affiliation{Universidad de Chile, Departamento de Astronom\'ia, Camino El Observatorio 1515, Las Condes, Santiago, Chile}

\author{N.~van~der~Marel}
\affiliation{National Research Council of Canada Herzberg Astronomy and Astrophysics Programs, 5071 West Saanich Road, Victoria, BC, V9E 2E7, Canada}

\begin{abstract}
We analyze the dust morphology of 29 transition disks (TDs) observed with Atacama Large (sub-)Millimeter Array (ALMA) at (sub-) millimeter emission. We perform the analysis in the visibility plane to characterize the total flux, cavity size, and shape of the ring-like structure. First, we found that  the $M_{\rm{dust}}-M_\star$ relation is much flatter for TDs than the observed trends from samples of class II sources in different star forming regions. This relation demonstrates that cavities open in high (dust) mass disks, independent of the stellar mass. The flatness of this relation contradicts the idea that TDs are a more evolved set of disks. Two potential reasons (not mutually exclusive) may explain this flat relation: the emission is optically thick or/and millimeter-sized particles are trapped in a pressure bump. Second, we discuss our results of the cavity size and ring width in the context of different physical processes for cavity formation. Photoevaporation is an unlikely leading mechanism for the origin of the cavity of any of the targets in the sample. Embedded giant planets or dead zones remain as potential explanations. Although both models predict correlations between the cavity size and the ring shape for different stellar and disk properties, we demonstrate that with the current resolution of the observations, it is difficult to obtain these correlations. Future observations with higher angular resolution observations of TDs with ALMA will help to discern between different potential origins of cavities in TDs. 

\end{abstract}


%%%%%%%%%%
\section{Introduction}     \label{sect:introduction}
%%%%%%%%%%
While planets are formed in the dense environments of protoplanetary disks, it is expected that the disk morphology evolves with time, creating a large diversity of structures. In this context, transition disks (TDs) are a terrific set of protoplanetary disks to witness the imprints of disk evolution. This set of disks was identified by the lack of near-infrared emission in the spectral energy distribution, which suggests an absence of material in the inner disk or the formation of a dust cavity \cite[or hole; e.g.,][]{strom1989, calvet2005, andrews2011, espaillat2014}. The origin of these cavities is still under debate, and accessing them through high angular resolution observations at different wavelengths has become indispensable to understand the formation of their structures and the processes of disk dispersal. Potential origins for TDs cavities include a nascent giant planet (or multiple planets) within the cavity, clearing up disk material \citep[e.g.][]{marsh1992, papaloizou2007, baruteau2014}, internal photoevaporation \cite[e.g.][]{alexander2007, owen2010, gorti2015, ercolano2017}, and regions of low disk ionization or dead zones \citep{regaly2012, flock2015, pinilla2016, ruge2016}.  However, it is idealistic to think that a single process is responsible for the diversity of the observed structures, as disk evolution can occur simultaneously through different mechanisms. 

Since the Atacama Large (sub-)Millimeter Array (ALMA) started its operations and released the first images in 2012, our understanding of the observed structures of TDs has been revolutionized. ALMA has confirmed that TDs are a very diverse set of protoplanetary disks, for which several gas and dust morphologies have been observed. Complementary to ALMA data, extreme adaptive optics and coronagraphic observations at optical and near-infrared wavelengths have also enriched our knowledge of the TDs structures \citep[e.g.][]{follette2013, avenhaus2014, deboer2016, pohl2017}. The combination of high-resolution observations at different wavelengths has showed how the distribution of micron-sized particles traced at short wavelengths can significantly differ from the distribution of millimeter-sized particles \citep[e.g.][]{garufi2013, pinilla2015a, hendler2018}. 

From the ALMA observations, different disks classified as TDs reveal themselves as not just a single cavity with only one surrounding ring-like emission, but instead they have several gaps and rings detected at the dust continuum emission. This is the case for the TDs around TW\,Hya, HD\,169142, HD\,97048 \citep[e.g.][]{andrews2016, walsh2016, fedele2017, vanderplas2017a}. In other cases, TDs have shown astonishing single high contrast asymmetries \citep[e.g HD\,142527 and IRS\,48,][]{casassus2013, vandermarel2013}, or more complex structures such as spiral structures  \citep[e.g. AB\,Aur,][]{tang2017}, or a combination of single rings and crescent structures, such as the TD around MWC\,758 and V1247\, Orionis \citep{kraus2017, boehler2017}. 

In total, ALMA has already observed dozens of TDs. It is thus timely to uniformly analyze this dataset and characterize the relationships between properties of these disks and their hosting stars. In this paper, we  analyze the dust morphology of a total of 29 TDs. Our sample starts with the TDs that have been observed from Cycle\,0 to Cycle\,3 (i.e. from 2012 to 2016) with an average resolution of $\sim20-40$\,au. To create a more uniform dataset, we exclude the disks for which multiple rings or gaps or strong azimuthal asymmetries have been detected at millimeter emission. The main objective of this paper is to characterize the cavity size and the radial shape of the ring-like emission in order to test theories of cavity formation and dust evolution. Therefore we focus our analysis on performing visibility modeling of only the real part, assuming that in our sample, disks are mainly symmetric. 

This paper is organized as follows. In the following section, we present a brief summary for the some of the potential origins of cavities in TDs and potential observational consequences. In Section~\ref{sect:observations}, we describe the observations and list the TDs selected in our sample. In Section~\ref{sect:analysis}, we present the analysis and results of quantifying the dust morphology of the TDs in this ALMA sample. In Section~\ref{sect:discussion}, we discuss our results in the context of the physical origin of the cavity and dust trapping. Finally, the conclusions of this work are summarized in Section~\ref{sect:conclusions}.

%%%%%%%%%%
\section{Potential origins of cavities in TDs}     \label{sect:origins}
%%%%%%%%%%

Some of the most common explanations for the origin of the cavities in TDs include planet-disk interactions, dead zones, and photoevaporation. For each of these physical processes, it is expected that the cavity size may depend on the stellar and disk properties. 

On one hand, if giant planets are responsible for the origin of the cavities in TDs, it is expected that cores of giant planets can form more efficiently around more massive stars \citep[e.g.][]{kennedy2008}, and in more massive disks \citep[e.g.][]{ida2004, mordasini2012}. 
Surveys of protoplanetary disks at millimeter wavelengths in different star forming regions show that the dust disk mass increases with the mass of the host star \citep[e.g.][]{andrews2013, ansdell2016, ansdell2017, barenfeld2016, pascucci2016, ward2018}. These results suggest that disks around more massive stars have more material to form more massive planets. 

From exoplanet surveys, it seems that giant planets are more frequent around more massive and more metal-rich stars \citep[e.g.][]{santos2004, udry2007, johnson2010}. However, for small planets, there is not a systematic correlation \citep[e.g.][]{mulders2015}. As a consequence, if cavities in TDs are due to giant planets, there might be a positive relation between stellar and disk mass and cavity size. Nonetheless, this picture becomes more complex due to the ability of super-Earth planets to open gaps or cavities in inviscid disks \citep[i.e. weakly turbulent disks; e.g.,][]{fung2017}. In addition, the ability of a planet with a given mass to carve a gap depends on location, as the relevant scales of the Hill radius and the scale height both vary with orbital distance -- and, in flared disks, not in proportion to each other \citep{crida2006, youdin2013, rosotti2016}. 

Dead zones can create structures as observed in TDs, because a gas density bump (and hence a pressure bump) can be formed at the outer edge of a dead zone as a result of the reduction of gas accretion in the dead zone. In this pressure bump, particles can grow and accumulate during million-year timescales, creating a large dust-cavity observable at different wavelengths \citep{pinilla2016}. In this case, \cite{dzyurkevich2013} demonstrated that the extension of the dead zone in the disk midplane depends on the stellar mass and disk mass (in particular, on the fraction of dust). From these models, it is expected that the outer edge of the dead zones is further out for more massive disks and around more massive stars. As a consequence, this scenario is expected also to predict a positive correlation between the cavity size and the disk and stellar mass. However, there are other important factors that can change the dead zone shape; for example, the magnetic field strength and the minimum grain size of the dust distribution \citep[e.g.][]{dzyurkevich2013}.

On the other hand, in the case of photoevaporation, the dependency of the cavity size with the stellar and disk properties is complex, since it depends on the stellar X-ray, EUV, and FUV radiation, for which there are not enough observational constraints to predict  cavity size as a function of stellar or disk mass. If the same ionizing photons clear disks around stars of different masses, the cavity size should scale with stellar mass \citep[e.g.][]{alexander2014}. The models of photoevaporation do have a clear prediction between cavity size and disk accretion rate, and predict that TDs with small cavities and with low accretion rates can be the product of photoevaporation \citep[e.g.][]{owen2012}. Recently, \cite{ercolano2018} demonstrated that photoevaporation can also create a large range of cavity sizes and accretion rates if the disk has moderate gas-phase depletion of carbon and oxygen around a Sun-like star (0.7\,$M_\odot$). 

In any of the three cases (planets, dead zones, or photoevaporation), millimeter-sized particles are expected to be trapped near to or further out of the edge of the cavity. The efficiency of the accumulation of dust particles in a pressure trap depends on the coupling of dust particles to the gas (i.e. on the disk mass), and also on the spatial location of the particle trap and the mass of the hosting star \citep[e.g.][]{nakagawa1986, brauer2008, pinilla2013}, as discussed in Sect~\ref{sect:model_predic}. 

%%%%%%%%%%%%
%FIGURE 
%%%%%%%%%%%%
\begin{figure*}
 \centering
   	\includegraphics[width=1.0\textwidth]{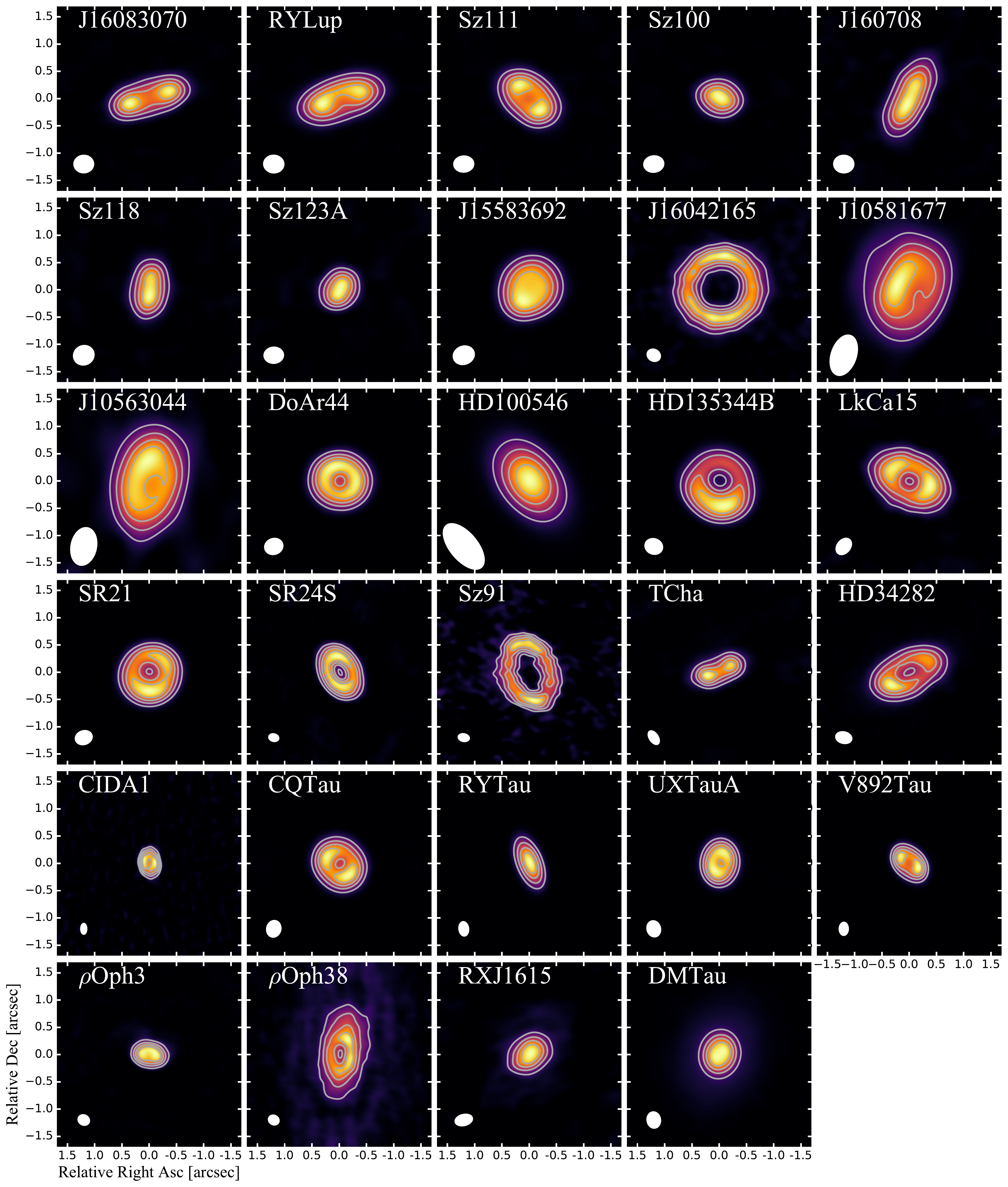}
	   \caption{ALMA dust continuum maps of the TDs considered in this work (Table~\ref{table:all_disks}). Contour lines at 20\%, 40\%, 60\%, 80\%, and 100\% value of the peak of emission of each target are over-plotted. The beam is shown for each case at the lower left part of each panel. }
   \label{fig:all_ALMA_maps}
\end{figure*}

%%%%%%%%%%%%
%TABLE 
%%%%%%%%%%%%
\begin{table*}
\caption{Disks Considered In This Work}
\label{table:all_disks}
\centering   
\begin{tabular}{|c|c|c|c|c|c|c|c|c|c|c|}
\hline
\hline       
\textbf{Target}&
R.A.&
Decl.&
beam&
$\log_{10}{M_\star}$&
$\log_{10}{\dot M_\star}$&
$i$&
PA&
{\scriptsize ALMA}&
REF\\
&
[J2000]&
[J2000]&
size [$\arcsec \times \arcsec$]&
[$M_\odot$]&
[$M_\odot$ y$^{-1}$]&
$[^{\circ}]$&
$[^{\circ}]$&
{\scriptsize Project} \#&
$(\dot M_\star)$\\
\hline
J16083070&16\,08\,30.68 & -38\,28\,27.22&$0.37\times0.33$&0.20&-8.98&74&107&{\scriptsize 2013.1.00220.S}&{\scriptsize \cite{alcala2017}}\\%&14-Jun-2015\\
\hline
RY\,Lup&15\,59\,28.37&-40\,21\,51.58&$0.38\times0.33$&0.14&-8.18&66&109&{\scriptsize 2013.1.00220.S}&{\scriptsize\cite{alcala2017}}\\%&14-Jun-2015 \\ 
\hline
Sz\,111&16\,08\,54.67&-39\,37\,43.49&$0.37\times0.30$&-0.32&-9.16&56&44&{\scriptsize 2013.1.00220.S}&{\scriptsize\cite{alcala2014}}\\%&15-Jun-2015\\
\hline
Sz\,100&16\,08\,25.75&-39\,06\,01.64&$0.38\times0.31$&-0.83&-9.45&45&60&{\scriptsize 2013.1.00220.S}&{\scriptsize\cite{alcala2014}}\\%&15-Jun-2015\\
\hline
J160708&16\,07\,08.54&-39\,14\,07.89&$0.38\times0.33$&-0.76&-9.20&73&155&{\scriptsize 2013.1.00220.S}&{\scriptsize\cite{alcala2017}}\\%&14-Jun-2015\\
\hline
Sz\,118&16\,09\,48.64&-39\,11\,17.24&$0.39\times0.36$&-0.04&-9.00&69&173&{\scriptsize 2013.1.00220.S}&{\scriptsize\cite{alcala2017}}\\%&14-Jun-2015\\
\hline
Sz\,123A&16\,10\,51.57&-38\,53\,14.10&$0.37\times0.30$&-0.29&-8.80&50&155&{\scriptsize 2013.1.00220.S}&{\scriptsize\cite{alcala2014}}\\%&15-Jun-2015\\
\hline
J15583692&15\,58\,36.90&-22\,57\,15.57&$0.40\times0.34$&0.05&$<$-11&30&148&{\scriptsize 2013.1.00395.S}&{\scriptsize\cite{rigliaco2015}}\\%&30-Jun-2014-\\
%&&&&&&&&&07-Jul-2014\\
\hline
J16042165&16\,04\,21.64&-21\,30\,28.98&$0.26\times0.22$&0.08&-10.54&6&80&{\scriptsize 2013.1.01020S}&{\scriptsize Pinilla et al. (in prep)}\\%&21-Jul-2015\\
\hline
J10581677&10\,58\,16.71&-77\,17\,17.15&$0.77\times0.47$&0.10&-7.81&66&160&{\scriptsize 2013.1.00437.S}&{\scriptsize\cite{manara2014}}\\%&01-May-2014-\\
%&&&&&&&&&03-May-2014\\
\hline
J10563044&10\,56\,30.31&-77\,11\,39.25&$0.71\times0.47$&-0.07&-9.41&39&160&{\scriptsize 2013.1.00437.S}&{\scriptsize\cite{manara2016}}\\%&01-May-2014\\
%&&&&&&&&&03-May-2014\\
\hline
DoAr\,44&16\,31\,33.46&-24\,27\,37.52&$0.35\times0.30$&0.11&-8.20&20&30&{\scriptsize 2012.1.00158.S}&{\scriptsize\cite{manara2014}}\\%&26-Jul-2014-\\
%&&&&&&&&&27-Jul-2014\\
\hline
HD\,100546&11\,33\,25.36&-70\,11\,41.27&$1.0\times0.51$&0.27&-7.04&44&146&{\scriptsize 2011.1.00863.S}&{\scriptsize\cite{fairlamb2015}}\\%&18-Nov-2012\\
\hline
HD\,135344B&15\,15\,48.42&-37\,09\,16.33&$0.34\times0.29$&0.17&-7.37&20&63&{\scriptsize 2012.1.00158.S}&{\scriptsize\cite{fairlamb2015}}\\%&27-Jul-2014-\\
%&&&&&&&&&28-Jul-2014\\
\hline
LkCa\,15&04\,39\,17\,80&+22\,21\,03.22&$0.34\times0.25$&0.00&-8.40&55&60&{\scriptsize 2011.0.00724.S}&{\scriptsize\cite{manara2014}}\\%&28-Aug-2012\\
\hline
SR\,21&16\,27\,10.27&-24\,19\,13.01&$0.32\times0.26$&0.29&-7.90&15&14&{\scriptsize 2012.1.00158.S}&{\scriptsize\cite{manara2014}}\\%&26-Jul-2014-\\
%&&&&&&&&&27-Jul-2014\\
\hline
SR\,24S&16\,26\,58.50&-24\,45\,37.20&$0.19\times0.15$&-0.09&-7.50&46&25&{\scriptsize 2013.1.00091.S}&{\scriptsize\cite{natta2006}}\\%&26-Sep-2015\\
\hline
Sz\,91&16\,07\,11.57&-39\,03\,47.85&$0.21\times0.15$&-0.31&-8.73&51&17&{\scriptsize 2013.1.00663.S}&{\scriptsize\cite{alcala2014}}\\%&11-Aug-2015\\
\hline
T\,Cha&11\,57\,13.28&-79\,21\,31.72&$0.29\times0.17$&0.03&-8.40&73&113&{\scriptsize 2012.1.00182.S}&{\scriptsize\cite{schisano2009}}\\%&17-Jul-2015\\
\hline
HD\,34282&05\,16\,00.48&-09\,48\,35.42&$0.31\times0.22$&0.27&$<$-8.30&60&118&{\scriptsize 2013.1.00658.S}&{\scriptsize\cite{fairlamb2015}}\\%&12-Dec-2014\\
%&&&&&&&&&31-Aug-2015\\
\hline
CIDA1&04\,14\,17.62&+28\,06\,0.9.28&$0.21\times0.12$&-0.96&-8.40&37&12&{\scriptsize 2015.1.00934.S}&{\scriptsize \cite{pinilla2018}}\\%&12-Aug-2016\\
\hline
CQ\,Tau&05\,35\,58.47&+24\,44\,53.70&$0.32\times0.27$&0.17&$<$-8.30&37&46&{\scriptsize 2013.1.00498.S}&{\scriptsize\cite{mendigutia2011}}\\%&30-Aug-2015\\
\hline
RY\,Tau&04\,21\,57.42&+28\,26\,35.13&$0.27\times0.19$&0.38&-7.20&62&23&{\scriptsize 2013.1.00498.S}&{\scriptsize\cite{cutri2003}}\\%&29-Sep-2015\\
\hline
UX\,TauA&04\,30\,04.00&+18\,13\,49.18&$0.31\times0.26$&0.14&-8.71&42&166&{\scriptsize2013.1.00498.S}&{\scriptsize\cite{rigliaco2015}}\\%&12-Aug-2015\\
\hline
V892\,Tau&04\,18\,40.62&+28\,19\,15.19&$0.25\times0.18$&0.45&---&55&52&{\scriptsize 2013.1.00498.S}&---\\%&29-Sep-2015\\
\hline
$\rho$Oph\,3&16\,23\,09.22&-24\,17\,05.36&$0.22\times0.19$&---&---&49&82&{\scriptsize 2013.1.00157.S}&---\\
\hline
$\rho$Oph\,38&16\,39\,45.73&-24\,02\,04.19&$0.21\times0.18$&---&---&0&45&{\scriptsize2013.1.00157.S}&---\\
\hline
RXJ1615&16\,15\,20.23&-32\,55\,05.36&$0.33\times0.21$&0.04&-8.50&45&153&{\scriptsize 2011.0.00724.S}&{\scriptsize\cite{manara2014}}\\%&14-Aug-2012\\
\hline
DM\,Tau&04\,33\,48.75&+18\,10\,09.66&$0.31\times0.25$&-0.27&-8.29&35&156&{\scriptsize 2013.1.00498.S}&{\scriptsize\cite{rigliaco2015}}\\%&12-Aug-2015\\
\hline
\hline
\end{tabular}    
\end{table*}

%%%%%%%%%%
\section{Observations and Sample of TDs}     \label{sect:observations}
%%%%%%%%%%

Our sample encompasses ALMA observations from Cycle\,0 to Cycle\,3. For most of the disks, the final measurement sets after self calibration were obtained from the principle investigator or co-investigators of the corresponding ALMA proposals. Otherwise, the data is taken as delivered from the ALMA archive, and there is no additional self-calibration performed (only one disk, T\,Cha). The sample includes the following disks: J16083070-3828268 (hereafter J16083070), RY\,Lup, Sz\,111, Sz\,100,  J16070854-3914075 (hereafter J160708), Sz\,118, and Sz\,123A from the most recent survey of the Lupus star-forming region \citep{ansdell2016}. From this region, there are other disks with tentative cavities \citep{tazzari2017a, vandermarel2018}, but they are excluded from the sample since the cavity size remains unconstrained from our visibilities analysis, as explained in Section~\ref{sect:analysis}. From the most recent ALMA survey of the Upper Sco star forming region, we perform our visibility analysis for all the potential TDs reported in \cite{barenfeld2016}, which may show a clear null at the interferometric visibilities as evidence for the existence of a cavity \citep[e.g.][]{hughes2007}. We performed the analysis for J15534211-2049282, J15583692-2257153, J16020757-2257467, J16042165-2130284, J16062196-1928445, J16063539-2516510, J16064102-2455489, and J16163345-2521505. Nonetheless, we only keep J15583692-2257153 and J16042165-2130284 (hereafter J15583692 and J16042165, respectively), for which the cavity size is well constrained. For J16042164, we took the most recent data in Band 6 with higher resolution and sensitivity \citep{dong2017}, and perform the final analysis with this data set.  From the ChaI star forming region \citep{pascucci2016}, we include two TDs, J10581677-7717170 and  J10563044-7711393 (hereafter J10581677 and J10563044, respectively), for which the cavity is resolved. Other TDs are included in our sample that have been observed individually, such as DoAr\,44 \citep{vandermarel2016}, HD\,100546 \citep{walsh2014}, HD\,135344B, SR\,21 \citep{pinilla2015b}, LkCa15, RXJ1615.3-3255 \citep{vandermarel2015}, SR\,24S \citep{pinilla2017}, Sz\,91\citep{canovas2016}, T\,Cha (ALMA project: 2012.1.00182.S, PI: J. Brown), and HD\,34282 \citep{vanderplas2017b}. In addition, we include from the Taurus star forming region the following disks: CIDA1 \citep{pinilla2018}, CQ\,Tau,  RY\,Tau, UX\,TauA, V892\,Tau, and DM\,Tau (all from ALMA project: 2013.1.00498.S, PI: L.\, Perez). Finally, we include two newly discovered TDs in the $\rho$ Ophiuchus molecular cloud: $\rho$Oph\,3 and $\rho$Oph\,38 \citep{cox2017}. Several disks of this sample have been re-observed in more recent years with higher resolution and sensitivity (e.g., HD\,100546), however, we prefer to use the data that has already been published to create a homogeneous sample at medium resolution.  

As mentioned above, we exclude TDs that show high contrast (higher than $\sim$2:1 contrast) or complex asymmetries at the continuum emission (e.g. IRS\,48, HD\,142527, MWC\,758, AB\,Aur), and TDs that now show multiple rings and cavities at the millimeter emission (e.g. TWHya, HD\,169142, HD\,97048). However, in our sample, we keep the TDs whose asymmetries are low contrast or that are debatable \citep[e.g. SR21, HD\,135344B, and HD\,34282,][]{pinilla2015b, vanderplas2017b}. In general, our interest is focused on large axisymmetric cavities. 

Since we aim to also test models of dust evolution, we analyze all disks in a similar wavelength (Band 7, i.e. $\sim0.87$\,mm; or Band 6, i.e. $\sim1.3$\,mm), to trace similar grain size.  Some of these TDs have been observed at multiple wavelengths with ALMA (e.g. SR\,24S or T\,Cha observed in Band 9/6 and Band 7/3, respectively), in which case we take either the Band 7 or Band 6 observations. Nonetheless, there are two disks for which we only have Band 9 observations; these are LkCa\,15 and  RXJ1615.3-3255. Although RXJ1615.3-3255 is excluded for our final analysis, as explained below, we keep LkCa\,15. 

The summary of the sample is found in Table~\ref{table:all_disks}, and the images after performing the {\tt clean} algorithm in the Common Astronomy Software Applications package \citep[CASA;][]{mcmullin2007} are shown in Fig.~\ref{fig:all_ALMA_maps}.  We used natural weighting for imaging \citep[except for CIDA1 that uses uniform weighting;][]{pinilla2018}, to obtain as high sensitivity as possible. The final representative angular resolution is reported in Table~\ref{table:all_disks}. Contour lines at 20\%, 40\%, 60\%, 80\%, and 100\% value of the peak of emission are over-plotted. The center, position angle (PA), and inclination ($i$) reported in Table 1 are found by fitting the data using {\tt uvmodelfit} in CASA, assuming a Gaussian and a disk model. In most of the cases, the Gaussian model provided lower uncertainties, and therefore we report the values using the Gaussian fit. In some cases, when the fit is poor, we performed the fit using only short baselines (baselines shorter than the location of the null in the real part of the visibilities), which guaranteed that the cavity is excluded from the fit. In general, limiting the baselines helps to decrease the uncertainties in the derivation of the disk center, inclination, and PA. In all cases, low uncertainties (within 5-8\% of the mean value) are found for the center, PA, and inclination, except for the disk around $\rho$Oph\,38, for which the inclination remains uncertain. The center reported in Table 1 is used to fix the visibilities using {\tt fixvis} in CASA, and these measurement sets are later used for imaging (Fig.~\ref{fig:all_ALMA_maps}). Such fixed visibilities are then deprojected using the PA and inclinations reported in Table~\ref{table:all_disks}. As explained in Sect.~\ref{sect:analysis}, we also performed some tests where the PA, inclination, and center remain as free parameters when fitting the visibilities. These tests provided similar results as using a priori {\tt uvmodelfit} and {\tt fixvis}. 

The stellar mass reported in Table~1 is calculated assuming the same evolutionary tracks for all the targets and employing the method described in \cite{pascucci2016}, which uses the evolutionary tracks from \cite{baraffe2015} and \cite{feiden2016}. However, for $\rho$Oph\,3 and $\rho$Oph\,38, we did not find information about the stellar luminosity and temperature in the literature, and we do not have stellar masses and accretion rates for these objects.  We update the distance of the targets that have been observed with Gaia \footnote{\url{https://www.cosmos.esa.int/gaia}} \citep[e.g. RY\,Lup, J15583692, HD 100546, HD\,135344B, T\,Cha, HD\,34282, RY\,Tau, and UXTauA;][]{gaia2016}. Finally, the corresponding citations for the accretion rates are in the last column of Table~\ref{table:all_disks}. 

Several of the targets in our sample have been observed in the optical and near-IR, including RY\,Lup \citep{langlois2018}, J16042165 \citep{pinilla2015a, canovas2017}, DoAr\,44 \citep{avenhaus2018}, HD\,100546 \citep{Garufi2016}, HD\,135344B \citep{stolker2016}, LkCa\,15 \citep{thalmann2016}, SR\,21 \citep{follette2013}, Sz\,91 \citep{Tsukagoshi2014}, T\,Cha \citep{pohl2017b}, RY\,Tau \citep{Takami2013}, UX\,TauA \citep{Tanii2012}, and RXJ1615 \citep{deboer2016}. In these cases,  azimuthal global asymmetries like those seen in the (sub-)mm emission are rarely observed, and even if they are (e.g. LkCa\,15 or HD\,100546), they are mostly due to the scattering phase function. Instead, other local asymmetric features are very common in these objects, including spiral arms, dips of emission or shadows, and arcs.

%%%%%%%%%%%%
\section{Data Analysis and Results} \label{sect:analysis}
%%%%%%%%%%%%

%%%%%%%%%%%%
%TABLE 
%%%%%%%%%%%%
\begin{table*}
\caption{Targets, assumed distance, observed frequency, results from MCMC fits, and optical depth at the peak of emission.}
\label{table:all_disks2}
\centering   
\begin{tabular}{|c|c|c|c|c|c|c|c|}
\hline
\hline       
\textbf{Target}&
Assumed&
$\nu$&
$r_{\rm{peak}}$&
$\sigma_{\rm{int}}$&
$\sigma_{\rm{ext}}$&
$F_{\rm{Total}}$&
$\tau_{\rm{peak}}$\\
&
distance [pc]&
[GHz]&
[au]&
[au]&
[au]&
[mJy]&
\\
\hline
J16083070&200&335.8&77.88$^{+3.69}_{-2.55}$&2.96$^{+3.17}_{-2.12}$&32.91$^{+1.54}_{-1.91}$&128.86$^{+1.08}_{-1.06}$&0.67\\
\hline
RY\,Lup&150&335.8&67.81$^{+1.44}_{-1.45}$&26.84$^{+1.48}_{-1.44}$&26.23$^{+0.96}_{-0.96}$&263.87$^{+1.25}_{-1.24}$&0.72\\ 
\hline
Sz\,111&200&335.8&56.54$^{+4.05}_{-3.08}$&3.95$^{+3.57}_{-2.77}$&31.35$^{+1.40}_{-1.88}$&176.71$^{+1.18}_{-1.21}$&0.48\\
\hline
Sz\,100&200&335.8&32.11$^{+3.04}_{-1.85}$&2.05$^{+2.72}_{-1.50}$&15.71$^{+1.13}_{-1.45}$&53.61$^{+0.37}_{-0.37}$&0.35\\
\hline
J160708&200&335.8&37.18$^{+9.26}_{-6.94}$&10.10$^{+12.16}_{-7.07}$&64.71$^{+3.32}_{-4.17}$&85.03$^{+1.26}_{-1.23}$&0.47\\
\hline
Sz\,118&200&335.8&57.60$^{+8.30}_{-8.66}$&12.40$^{+7.12}_{-7.60}$&19.60$^{+4.20}_{-4.92}$&59.71$^{+0.63}_{-0.63}$&0.43\\
\hline
Sz\,123A&200&335.8&57.96$^{+2.36}_{-3.00}$&26.68$^{+4.73}_{-3.77}$&2.72$^{+2.48}_{-1.89}$&39.65$^{+0.62}_{-0.64}$&0.32\\
\hline
J15583692&166&341.1&84.07$^{+0.56}_{-1.14}$&66.15$^{+1.18}_{-1.22}$&0.78$^{+0.92}_{-0.50}$&175.30$^{+0.35}_{-0.36}$&0.36\\
\hline
J16042165&145&234.0&80.15$^{+1.45}_{-1.49}$&11.20$^{+1.27}_{-1.32}$&20.86$^{+1.01}_{-1.03}$&69.07$^{+0.80}_{-0.80}$&0.21\\
\hline
J10581677&180&338.0&67.56$^{+2.09}_{-1.61}$&1.24$^{+1.84}_{-0.90}$&73.97$^{+1.45}_{-1.49}$&329.96$^{+2.51}_{-2.62}$&0.50\\
\hline
J10563044&180&338.0&55.63$^{+4.01}_{-2.68}$&2.37$^{+3.54}_{-1.76}$&54.83$^{+2.39}_{-2.60}$&141.85$^{+2.93}_{-2.96}$&0.25\\
\hline
DoAr\,44&120&335.6&34.26$^{+0.21}_{-0.21}$&5.81$^{+0.18}_{-0.18}$&13.18$^{+0.12}_{-0.12}$&180.40$^{+0.20}_{-0.20}$&0.36\\
\hline
HD\,100546&109&346.3&14.89$^{+0.52}_{-0.51}$&0.25$^{+0.47}_{-0.19}$&19.14$^{+0.24}_{-0.27}$&1135.32$^{+2.04}_{-2.03}$&1.25\\
\hline
HD\,135344B&156&346.3&62.73$^{+0.10}_{-0.10}$&22.04$^{+0.10}_{-0.10}$&28.88$^{+0.06}_{-0.06}$&606.89$^{+0.41}_{-0.41}$&0.63\\
\hline
LkCa\,15&140&688.7&47.92$^{+0.98}_{-1.00}$&10.63$^{+0.92}_{-0.96}$&41.50$^{+0.58}_{-0.58}$&1458.14$^{+8.58}_{-8.35}$&0.91\\
\hline
SR\,21&120&346.3&50.94$^{+0.09}_{-0.09}$&19.60$^{+0.08}_{-0.08}$&6.94$^{+0.07}_{-0.07}$&347.03$^{+0.19}_{-0.19}$&0.49\\
\hline
SR\,24S&137&234.0&41.67$^{+0.06}_{-0.06}$&11.72$^{+0.06}_{-0.06}$&18.75$^{+0.04}_{-0.04}$&227.18$^{+0.15}_{-0.14}$&0.71\\
\hline
Sz\,91&200&338.2&96.19$^{+6.82}_{-4.95}$&5.01$^{+5.71}_{-3.57}$&37.01$^{+4.82}_{-5.32}$&34.33$^{+2.13}_{-2.03}$&0.23\\
\hline
T\,Cha&108&338.1&26.79$^{+0.16}_{-0.16}$&6.93$^{+0.15}_{-0.16}$&17.29$^{+0.09}_{-0.09}$&225.21$^{+0.22}_{-0.22}$&2.30\\
\hline
HD\,34282&325&351.3&138.97$^{+0.51}_{-0.51}$&35.21$^{+0.48}_{-0.48}$&57.30$^{+0.35}_{-0.36}$&333.67$^{+0.61}_{-0.60}$&0.79\\
\hline
CIDA1&140&338.1&28.76$^{+0.53}_{-0.86}$&12.66$^{+0.81}_{-0.82}$&0.57$^{+0.67}_{-0.41}$&35.40$^{+0.15}_{-0.15}$&0.40\\
\hline
CQ\,Tau&160&223.7&46.47$^{+0.17}_{-0.18}$&10.80$^{+0.15}_{-0.16}$&16.61$^{+0.11}_{-0.11}$&172.17$^{+0.14}_{-0.14}$&0.53\\
\hline
RY\,Tau&176&223.7&21.25$^{+0.29}_{-0.29}$&92.60$^{+5.38}_{-9.79}$&32.99$^{+0.15}_{-0.15}$&232.45$^{+0.27}_{-0.27}$&$\gg1$\\
\hline
UX\,TauA&158&223.7&37.53$^{+0.81}_{-0.91}$&4.48$^{+0.62}_{-0.71}$&7.73$^{+0.48}_{-0.45}$&64.94$^{+0.08}_{-0.08}$&0.39\\
\hline
V892\,Tau&140&223.7&33.51$^{+0.11}_{-0.11}$&8.69$^{+0.09}_{-0.09}$&9.21$^{+0.07}_{-0.07}$&286.71$^{+0.18}_{-0.18}$&$\gg1$\\
\hline
$\rho$Oph\,3&137&223.7&27.74$^{+0.89}_{-0.85}$&9.47$^{+0.79}_{-0.76}$&6.21$^{+0.56}_{-0.60}$&96.42$^{+0.35}_{-0.36}$&0.61\\
\hline
\hline
\end{tabular}   
\tablecomments{The optical depth at the peak of emission $\tau_{\rm{peak}}$ is calculated assuming a physical temperature of 20\,K throughout the disk for all the targets.} 
\end{table*}

To fit the millimeter dust continuum emission of all the disks, we homogeneously perform an analysis in the visibility domain. Since the disks in our sample appear axisymmetric at the resolution of the considered observations, we concentrated on fitting the real part of the visibilities (the imaginary part is identically zero or oscillates very close to zero after centering the sources). Our model consists of a radially asymmetric Gaussian ring for the millimeter intensity ($I(r)$)-that is, a Gaussian ring whose inner and outer widths ($\sigma_{\rm{int}}$ and $\sigma_{\rm{ext}}$, respectively) can differ, such that

\begin{equation}
I(r)=\left\{ \begin{array}{rcl}
C\exp\left(-\frac{(r-r_{\rm{peak}})^2}{2\sigma_{\rm{int}}^2}\right) &\mbox{for} & r\leq r_{\rm{peak}}\\
C\exp\left(-\frac{(r-r_{\rm{peak}})^2}{2\sigma_{\rm{ext}}^2}\right) &\mbox{for} & r> r_{\rm{peak}},
\end{array}\right.
\label{eq:asymmetric_model}
\end{equation}

\noindent where  $C$ is connected with the total flux of the disk as explained below. This profile was introduced in \cite{pinilla2017} to fit the morphology of TDs and to mimic the effect of particle trapping in a radial pressure bump. From dust evolution models, it is expected that under the presence of a single pressure bump, the external width of the ring is larger than the internal because in the  outer disk the particles take longer times to grow and drift toward the pressure maximum, creating a ring with an outer tail as discussed in Section~\ref{sect:model_predic}. 

%%%%%%%%%%%%
%FIGURE 
%%%%%%%%%%%%
\begin{figure}
 \centering
   	\includegraphics[width=1.0\columnwidth]{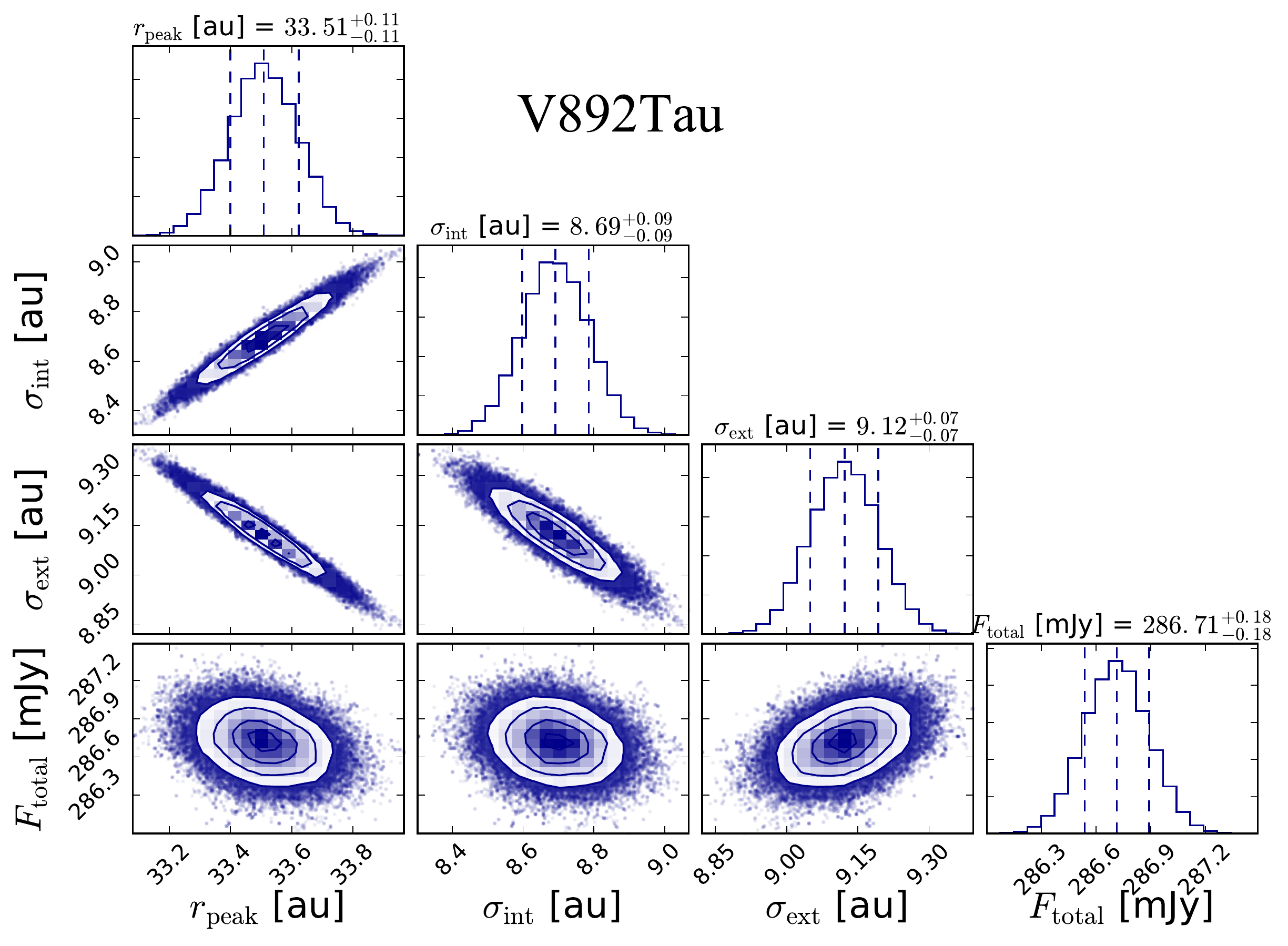}
	   \caption{MCMC results for V892\,Tau, showing the one-dimensional and two-dimensional posterior
distributions for the MCMC fit. The plot shows the posterior sampling provided by the last 700 steps of the 200 walkers chain. The median values and the 1$\sigma$  standard deviation of the best-fitting parameters are shown in vertical dashed lines.}
   \label{MCMC_fit}
\end{figure}

%%%%%%%%%%%%
%FIGURE 
%%%%%%%%%%%%
\begin{figure*}
 \centering
   	\includegraphics[width=1.0\textwidth]{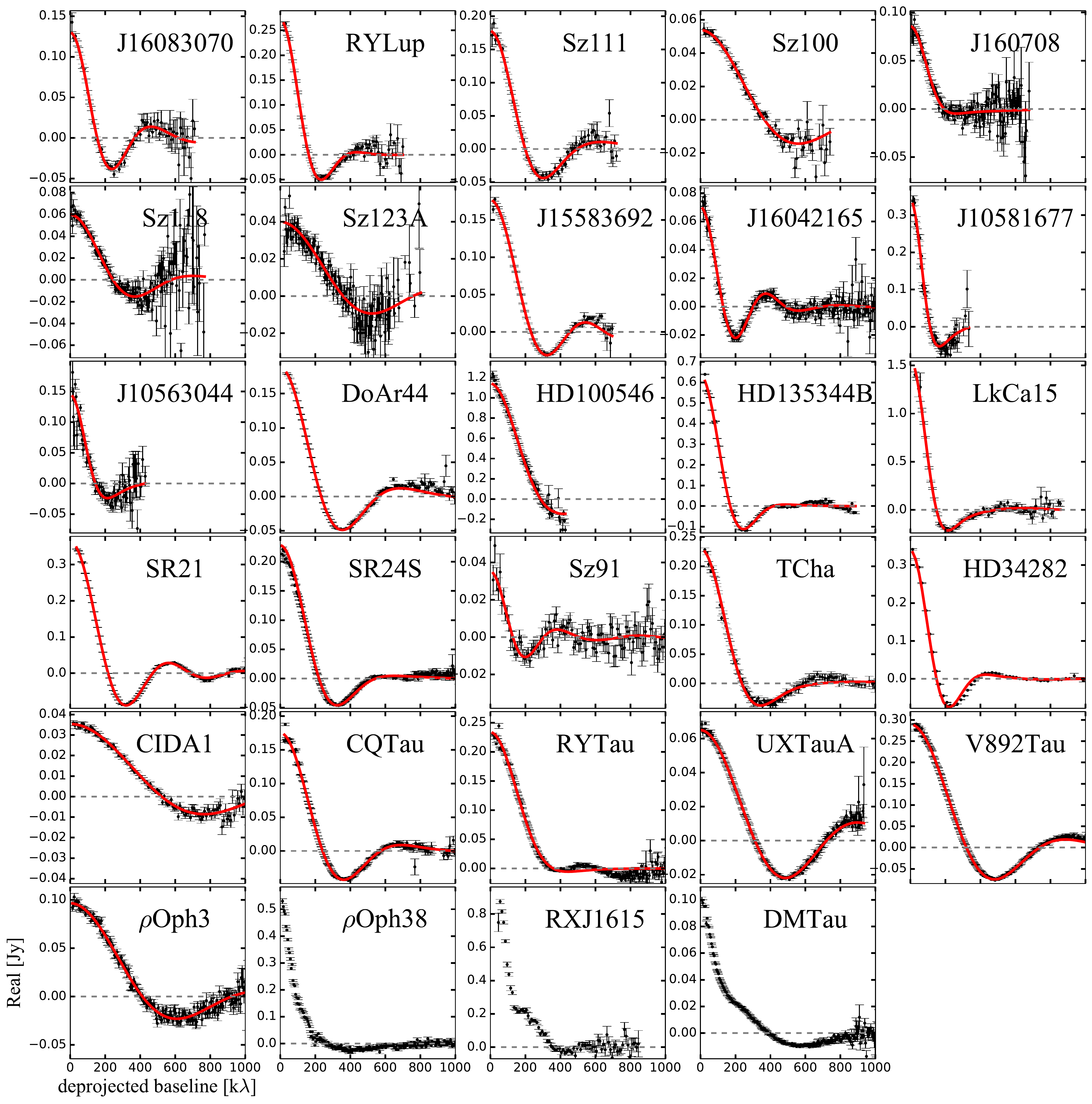}
	   \caption{Real part of the binned and deprojected visibilities for each target and the model with the best-fitting parameters (red solid lines) from the MCMC fit (Table~\ref{table:all_disks2}). The error bars correspond to the standard error in each bin.}
   \label{all_data_models}
\end{figure*}

For each disk, we deprojected the data to perform a fit in the visibility plane. The Fourier transform of an azimuthally symmetric brightness distribution can be expressed in terms of the zeroth-order Bessel function of the first kind $J_0$ of the deprojected uv-distance-$r_{uv}$ \citep{berger2007}

\begin{equation}
V_{\rm{Real}} (r_{uv})=2\pi\int^\infty_0 I(r) J_0(2\pi r_{uv}r)r dr, 
\label{eq:real_part}
\end{equation}
 
\noindent and therefore the constant $C$ of Eq.~\ref{eq:asymmetric_model} is related with the total flux as

\begin{equation}
C=\frac{F_{\rm{total}}}{\int^\infty_0 I(r) J_0(0)r dr}. 
\end{equation}

The fitting is performed using the Markov chain Monte Carlo (MCMC) method, and we used {\it emcee} \citep{foreman2013}, and follow the same procedure as in \cite{pinilla2017}. We explored four free parameters ($r_{\rm{peak}}$, $\sigma_{\rm{int}}$, $\sigma_{\rm{ext}}$, and $F_{\rm{total}}$) with 200 walkers and 1000 steps in each case, while the center, PA, and inclination are fixed. We adopted a set of uniform prior probability distributions for the free parameters explored by the Markov chain, such that

\begin{eqnarray}
r_{\rm{peak}} &\in& [1, 150]\,\rm{au} \nonumber\\
\sigma_{\rm{int}} &\in& [1, 100]\,\rm{au}\nonumber\\
\sigma_{\rm{ext}}&\in& [1, 100]\,\rm{au} \nonumber\\
F_{\rm{total}}&\in& [0.0, 2.0]\,\rm{Jy}
\label{eq:parameter_space}
\end{eqnarray}

As tests, we randomly chose three targets of our sample (J15583692, DoAr\,44, and SR\,21) and perform the fit of the visibilities leaving the inclination, PA, and center as free parameters, and using the publicly available code GALARIO \citep{tazzari2018}. These tests gave similar results in all three cases,  providing confidence about the  accuracy of our procedure. The radial grid in the model for the MCMC fit is taken linear, specifically $r\in[0-500]\,$au with steps of 0.5\,au, which is much lower than the observation's synthesized beam.  In most of the cases, the autocorrelation time of the MCMC fit is around 100 steps, and we take the last 700 steps to obtain the posterior distributions, the median values, and the $1\sigma$ standard deviation of the best-fitting parameters. The results of the MCMC fits are summarized in Table~\ref{table:all_disks2}, and Fig.~\ref{MCMC_fit} shows an example of the results of one of the fits (in this case for the TD around V892Tau). 

Figure~\ref{all_data_models} shows the binned data corresponding to the real part of the visibilities for each case, and we over-plot the model with the best-fitting parameters. The error bars correspond to the standard error in each bin. In addition, we checked the residuals (models-observations) in the visibility plane, which are always around zero within the uncertainty of the data. 

The last three targets ($\rho$Oph38, RXJ1615, and DMTau) are not fitted with this procedure, since the visibilities and the residuals evidence the existence of more than one ring. The visibilities of DM\,Tau were fitted by \cite{zhang2016}, who used two rings of emission to fit the visibilities at 329\,GHz. Our data at 223.7\,GHz show a bump of emission at around 200\,k$\lambda$ (deprojected baseline), which indicates the existence of a ring of emission further out of the cavity. Similarly, RXJ1615 shows a bump at similar position in the visibilities. This disk reveals several rings and gaps in the NIR scattered light emission \citep{deboer2016}, and these visibilities suggest the existence of substructures at millimeter emission around 200\,k$\lambda$. \cite{vandermarel2015} hinted at the presence of a dust gap in the outer part of the disk between 110 and 130\,au from the ALMA resulting image of the same dataset.  The disk around $\rho$Oph38 has a two-ringed-like structure as described in \cite{cox2017}. For the analysis and discussion in the following sections, we do not consider these three disks because we only keep TDs with a cavity and a single ring of emission. It remains part of future work to quantify the shape of such substructures in these three disks from visibility analysis, and high angular resolution observations are required to confirm these structures, as suggested by the visibilities. 

%%%%%%%%%%%%
%FIGURE 
%%%%%%%%%%%%
\begin{figure*}
 \centering
   	\includegraphics[width=1.0\textwidth]{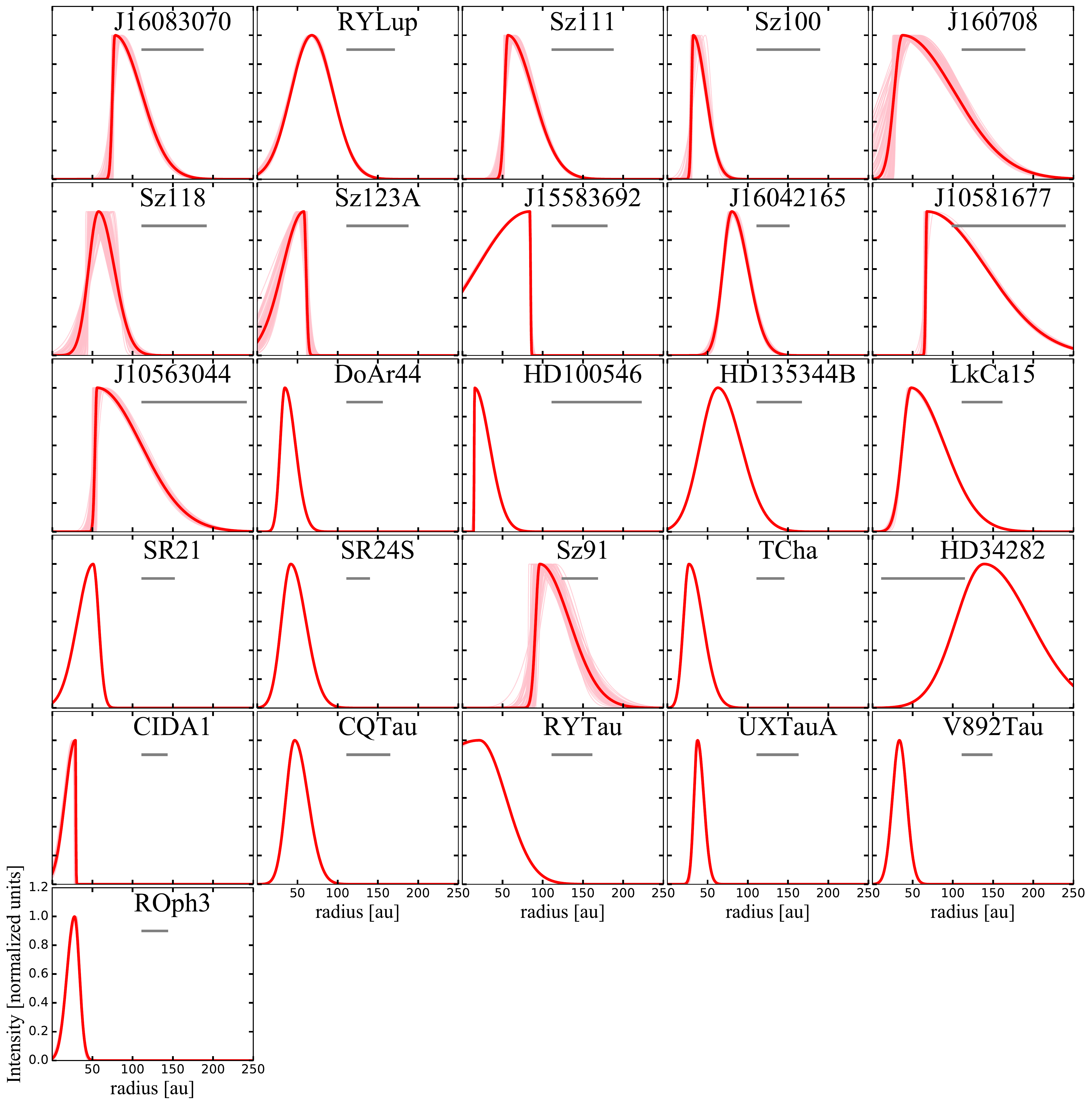}
	   \caption{Best fit models from the MCMC fits (Table~\ref{table:all_disks2}), assuming an intensity profile as Eq.~\ref{eq:asymmetric_model}. For each case, the intensity is normalized to the value at the location of peak of the ring.  We over-plot 100 models that randomly take a set of parameters from each sample of the MCMC fit. The horizontal gray line in each panel corresponds to the beam major axis. }
   \label{all_intensity_models}
\end{figure*}

Figure~\ref{all_intensity_models} shows the normalized intensity profile of the model, taking the median values of the best-fitting parameters. In addition, to give a diagnostic of  the convergency of the fit, we over-plot 100 models that randomly take a set of parameters from each sample.  In most of the cases, the best fit model shows a clear cavity and a ring-like emission, except for the disk around RY\,Tau. This disk was identified as a TD by imaging a cavity with CARMA at  1.3\,mm \citep{isella2010}, but the current ALMA observations at the same wavelength do not show a clear cavity. For this target, our MCMC model poorly fits the data between $\sim$400 and 600\,k$\lambda$, which appears in the residuals as a ring of emission at $\sim$0.$\arcsec$5 (or 88\,au assuming a distance of 176\,pc). This structure is more apparent in the most recent ALMA Cycle 4 data  (PI: G. Herczeg), in which at least two rings are required to fit the visibility data at 1.3\,mm (Long et al., in prep). Therefore, RY\,Tau is possibly in the category of disks with multiple rings and gaps, which is likely the reason for our poor fit.

The intensity profiles shown in Fig.~\ref{all_intensity_models} demonstrate the large diversity of cavity sizes and ring-like emission around the cavity, in some cases showing almost a perfectly symmetric radial ring, and in other cases where the inner or the outer edge of the ring is very truncated.  In the context of our models, $r_{\rm{peak}}$ corresponds to the cavity size when comparing with  observational analysis by for example, \cite{andrews2011} and \cite{vandermarel2016, vandermarel2018}, who fit the dust morphology by assuming a sharp edge of the dust density at the cavity location. The second motivation to use $r_{\rm{peak}}$ as the cavity size is that the models of dust trapping predict well the location of the pressure maximum (the peak of the millimeter emission), and this has been used to infer planet properties such as planet mass and position \citep[e.g.][]{dejuanovelar2013}.

%%%%%%%%%%%%%%%%%%%%%%%%%%%%%%%%%%%%
\section{Discussion} \label{sect:discussion}
%%%%%%%%%%%%%%%%%%%%%%%%%%%%%%%%%%%%

 %%%%%%%%%%%%
%FIGURE 
%%%%%%%%%%%%
\begin{figure}
 \centering
	\includegraphics[width=1.0\columnwidth]{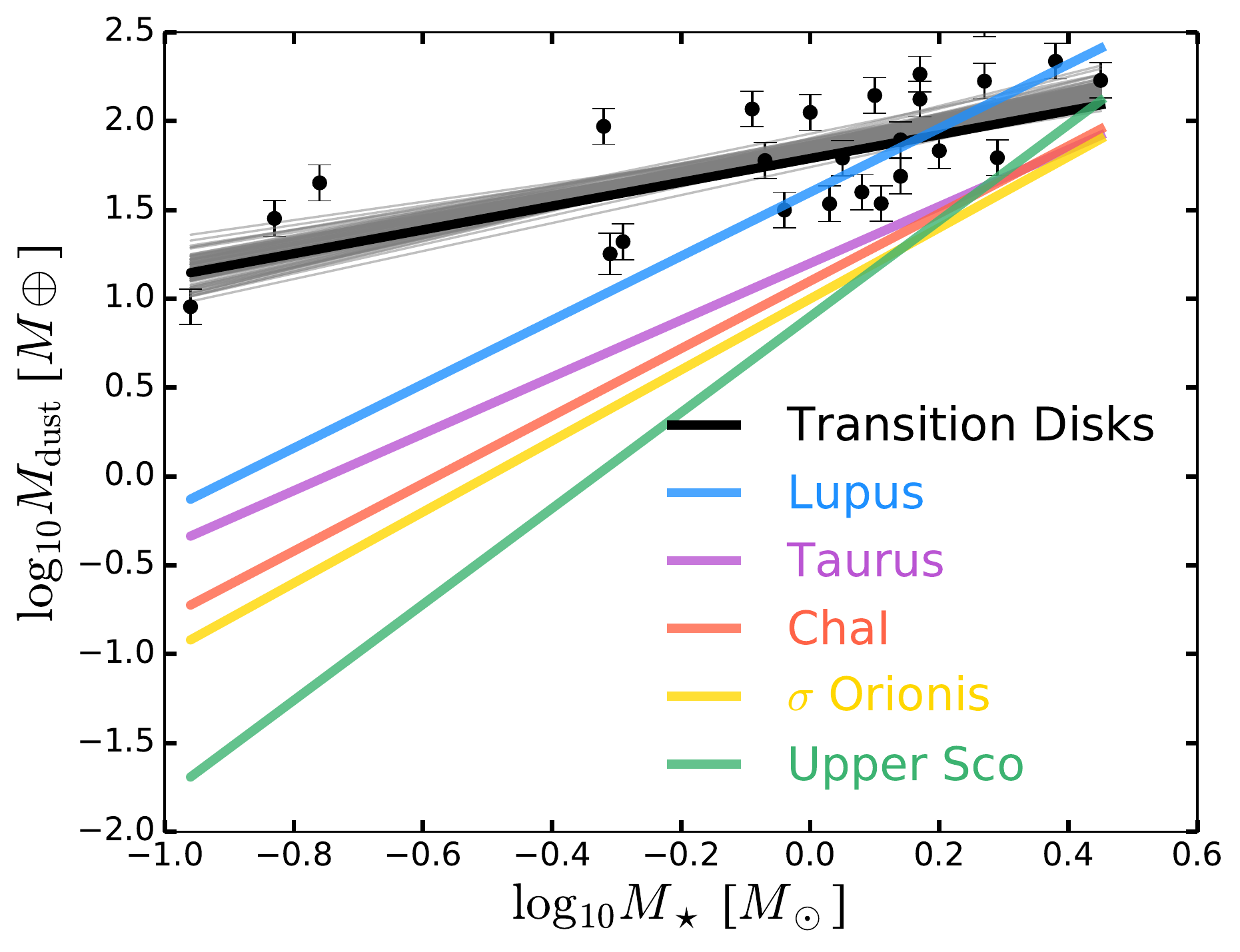}
	\caption{$M_{\rm{dust}}-M_{\star}$ relation in different star forming regions (colors) and for the TDs (black points) of the sample of this work. The values for the slope and intersect are taken from \cite{pascucci2016} (except for $\sigma$-Orionis), who used the same evolutionary tracks and constant temperature of 20\,K. For $\sigma$-Orionis, we took the values from \cite{ansdell2017}. From fitting this relation for TDs, we used $\log_{10}(M_{\rm{dust}}/M_\oplus)=\beta \log_{10}(M_\star/M_\odot)+\alpha$, and we obtained $\beta=0.72^{+0.08}_{-0.08}$  and $\alpha=1.85^{+0.03}_{-0.03}$. The fit takes into account the uncertainties of the data. For the uncertainties, we include 10\% of uncertainty from flux calibration for every source.}
   \label{Mdisk_Mstar_relation}
\end{figure}

\subsection{$M_{\rm{dust}}-M_\star$ Relation}

With the total flux obtained from the MCMC fit of each disk, we calculate the disk dust mass assuming optically thin emission, as in \cite{hildebrand1983}, 

\begin{equation}
	M_{\mathrm{dust}}\simeq\frac{{d^2 F_\nu}}{\kappa_\nu B_\nu (T(r))}
  \label{mm_dust_mass}
\end{equation}

\noindent where $d$ is the distance to the source and \ $\kappa_\nu$ is the mass absorption coefficient at a given frequency, which we assume to be  $\kappa_\nu=2.3\,$cm$^{2}$\,g$^{-1}\times(\nu/230\,\rm{GHz})^{0.4}$ \citep[][]{andrews2013}. $B_\nu (T_{\rm{dust}})$ is the Planck function for a given dust temperature $T_{\rm{dust}}$, for which we assume 20\,K in all cases. We consider the total fluxes obtained from the visibility fitting (i.e., the values reported in Table~\ref{table:all_disks2}).

Figure~\ref{Mdisk_Mstar_relation} shows the $M_{\rm{dust}}-M_\star$ relation for TDs in black color. We do not use color points for each TDs according to the star-forming region, because some of these targets are isolated. We fit a linear relation to these data (i.e. $\log_{10}(M_{\rm{dust}}/M_\oplus)=\beta \log_{10}(M_\star/M_\odot)+\alpha$) using an MCMC fit that takes into account the uncertainties of the data. For the uncertainties, we include 10\% of uncertainty from flux calibration for every source.  In the visibility fitting, it is possible to recover a slightly higher total flux than in the image plane. This is the case of, for example, the two TDs in ChaI when comparing with the values reported in \cite{pascucci2016}. We use the values of the total flux from the MCMC fit, which results in slightly higher dust disk masses too. From the fit, we obtained $\beta=0.72^{+0.08}_{-0.08}$ and $\alpha=1.85^{+0.03}_{-0.03}$, where the uncertainties are based on the 16th and 84th percentiles of the posterior distribution. Our sample of TDs spans different stellar ages ($\sim1-10$\,Myr), and the three disks that we have around very low mass stars are in the youngest star forming regions (i.e., Taurus and Lupus). 

We compare this relation with the previous fits obtained in different star-forming regions (Fig~\ref{Mdisk_Mstar_relation}). In particular, with the values reported in \cite{pascucci2016}, who performed the fit assuming for all the cases a constant temperature of 20\,K and used the same evolutionary tracks  and $\kappa_\nu$ as in our case. However, for $\sigma$-Orionis, we took the values reported by \cite{ansdell2017}, who also performed this fit for different star forming regions, but using the evolutionary tracks from \cite{siess2000}, which do not cover the low mass stars but lead to similar values as \cite{baraffe2015} for $>0.1\,M_\odot$. It is important to note that we also checked the results assuming that the dust disk temperature scales with stellar luminosity as in \cite{pascucci2016}, taking the $T_{\rm{dust}}-L_\star$ as in \cite{andrews2013}, and we found that in general  the trends remain similar. Nevertheless, for the rest of the paper we keep 20\,K temperature for all the sources, to avoid introducing an artificial bias in the masses that would have been reflected in the correlations that we investigate in Sect.~\ref{sect:cav_correlations}.

Previously, it was found that there is a steepening of the $M_{\rm dust}$-$M_{\star}$ relation with age \citep[e.g.][]{ansdell2016, ansdell2017, pascucci2016}, and in our case, this relation is much flatter for the TDs of our sample than for any other region. If TDs are a more evolved set of disks, our results seemingly
contradict our expectations. This relation also shows that cavities open in high (dust) mass disks, independent of the stellar mass.

There are two possible reasons for this flatter relation. First, it is possible that the millimeter emission is optically thick, and the dust disk mass is underestimated when using Eq.~\ref{mm_dust_mass}. This can be the case  when dust accumulates in particular disk regions, increasing the local dust-to-gas ratios significantly. If this preferentially affects the most massive disks, this will flatter the relation. To have an estimation of the optical thickness, we calculate the optical depth at the peak of emission ($\tau_{\rm{peak}}=-\ln[1-T_{\rm{brightness}}/T_{\rm{physical}}]$, being $T_{\rm{brightness}}$ and $T_{\rm{physical}}$ the brightness and physical temperature, respectively), assuming a physical temperature of 20\,K for all targets, and the values are reported in the last column of Table~\ref{table:all_disks2}. The brightness temperature is calculated from the blackbody Planck function without assuming the Rayleigh-Jeans regime, and taking the flux at the peak of emission. Adopting a constant temperature throughout the disk, the optical depth increases within the ring of emission until it reaches  $\tau_{\rm{peak}}$ and then it decreases outwards. The emission \emph{at the peak} of the ring is partially optically thick ($\tau_{\rm{peak}}\sim$ 0.2-1.0) in most cases and optically thick $\tau_{\rm{peak}}>1$ for four targets. 

The second potential reason is that the ring-like emission observed in TDs is indeed the result of particle trapping in pressure maximum. In pressure maxima, the radial drift of the millimeter- or centimeter-sized particles in the outer parts of the disks is completely suppressed or reduced. \cite{pascucci2016} demonstrated that to recover the  steepness of the $M_{\rm dust}$-$M_{\star}$ relation, dust evolution models that include the growth, fragmentation, \emph{and} drift of particles are needed. The steepness of the relation is only reproduced when radial drift is included  because it is expected to reduce the dust mass with time, and because drift is more effective around low mass stars \citep{pinilla2013}. However, if only growth and fragmentation of particles happens in a radial pressure bump (because drift is reduced or suppressed), the $M_{\rm dust}$-$M_{\star}$ relation is expected to be flatter \citep[see Fig. 9 in][]{pascucci2016}. As a consequence, it is possible that this very flat relation evidences that the structures seen in TDs are in fact the result of particle trapping.  

Motivated by the potential fact that the flatness of the relation seen in Fig.~\ref{Mdisk_Mstar_relation} is due to particle trapping, we discuss in the next sub-section the model predictions of particle trapping by, in particular, giant planets, and  explain how the cavity size and ring-like emission depend on stellar and disk properties in order to compare to our MCMC results. 

%%%%%%%%%%%%
%FIGURE 
%%%%%%%%%%%%
\begin{figure}
 \centering
   	\includegraphics[width=1.0\columnwidth]{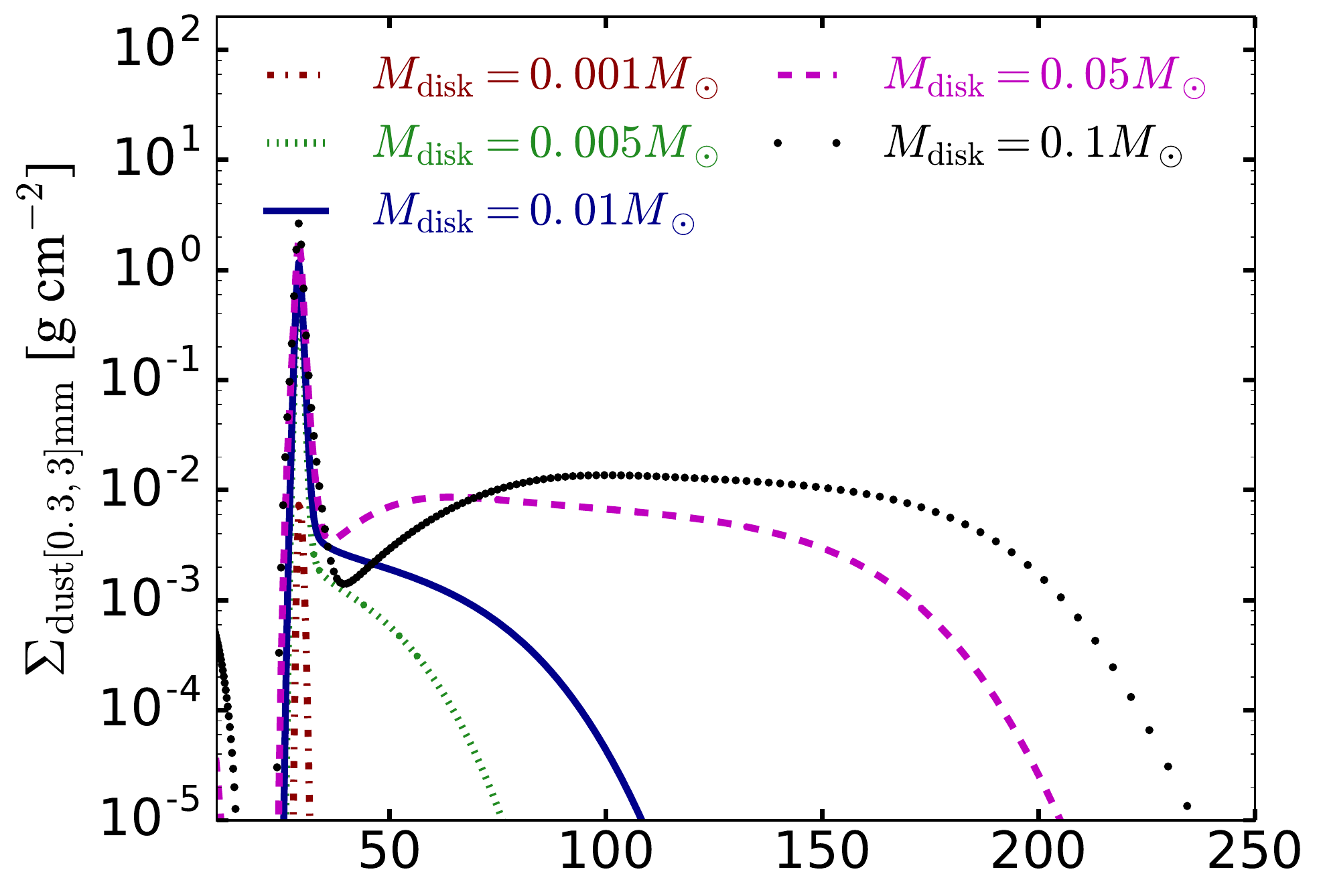}\\
	\includegraphics[width=1.0\columnwidth]{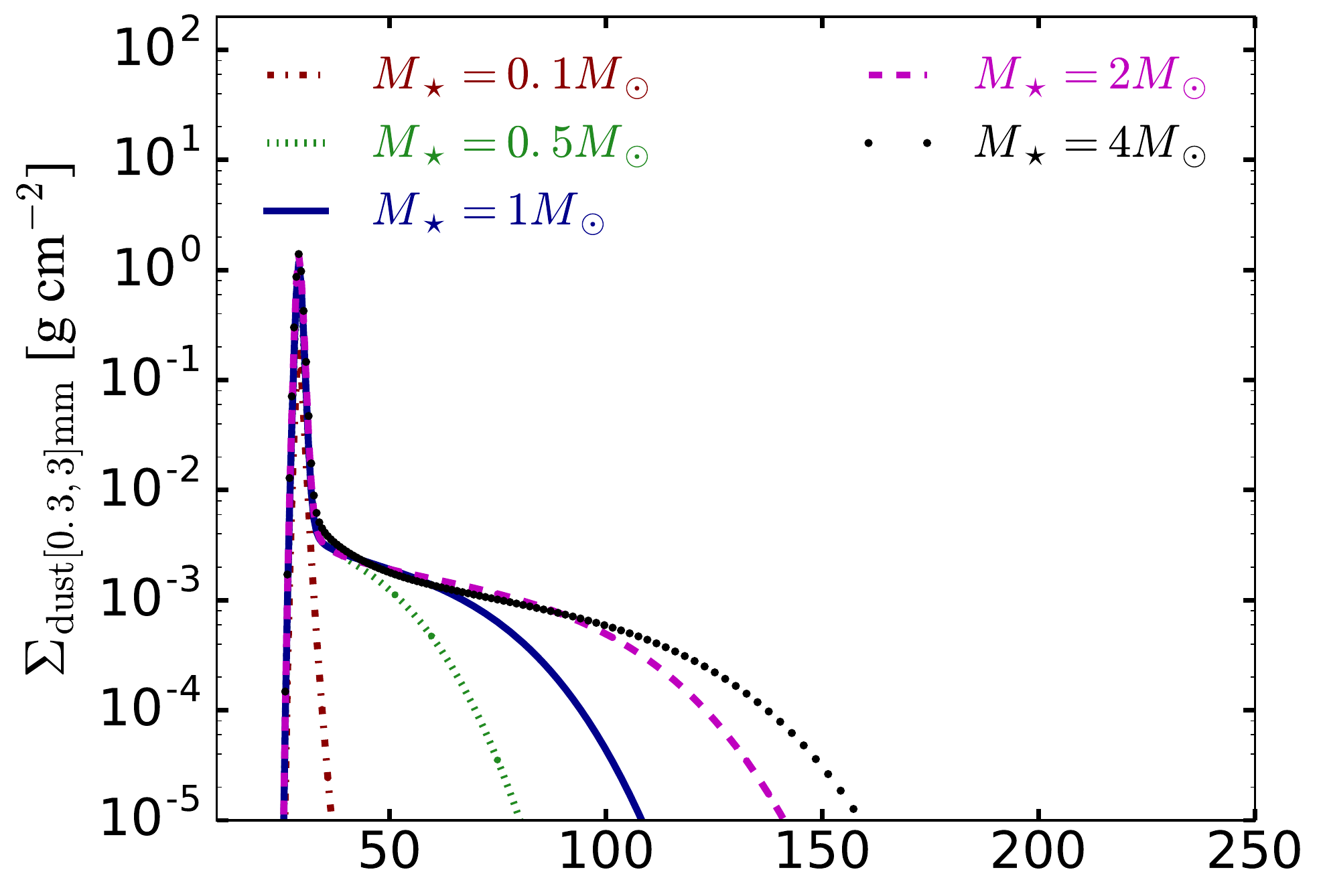}\\
	\includegraphics[width=1.0\columnwidth]{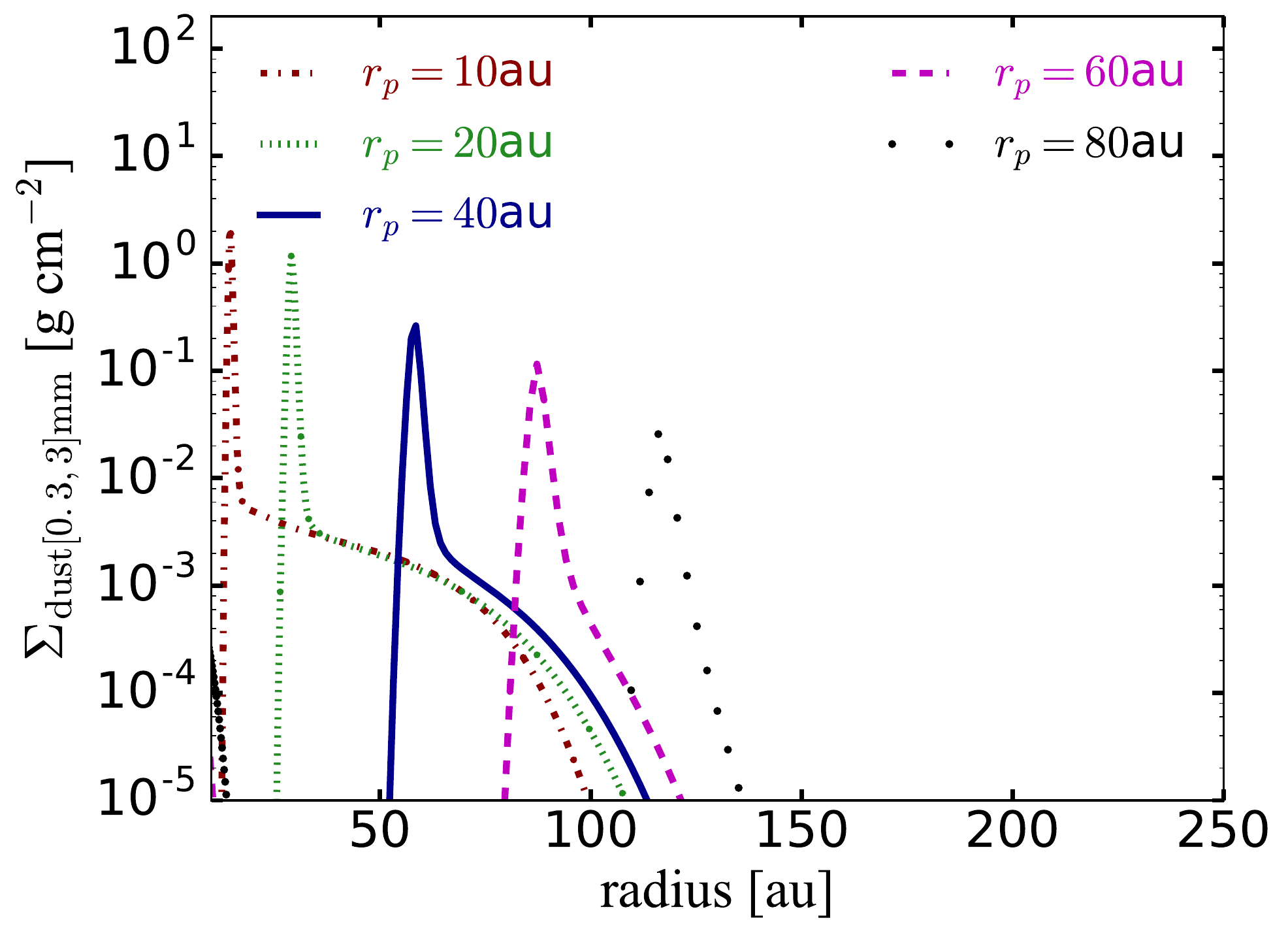}
	   \caption{Dust density distribution for particle sizes between 0.3 and 3\,mm after 1\,Myr of evolution, in the case where one Jupiter mass planet is embedded in the disk, and where different disk mass, stellar mass, or planet position are assumed.}
   \label{model_predictions}
\end{figure}

%%%
\subsection{Model predictions of trapping by giant planets} \label{sect:model_predic}
%%%

%%%%%%%%%%%%
%FIGURE 
%%%%%%%%%%%%
\begin{figure*}
 \centering
   	\includegraphics[width=1.0\textwidth]{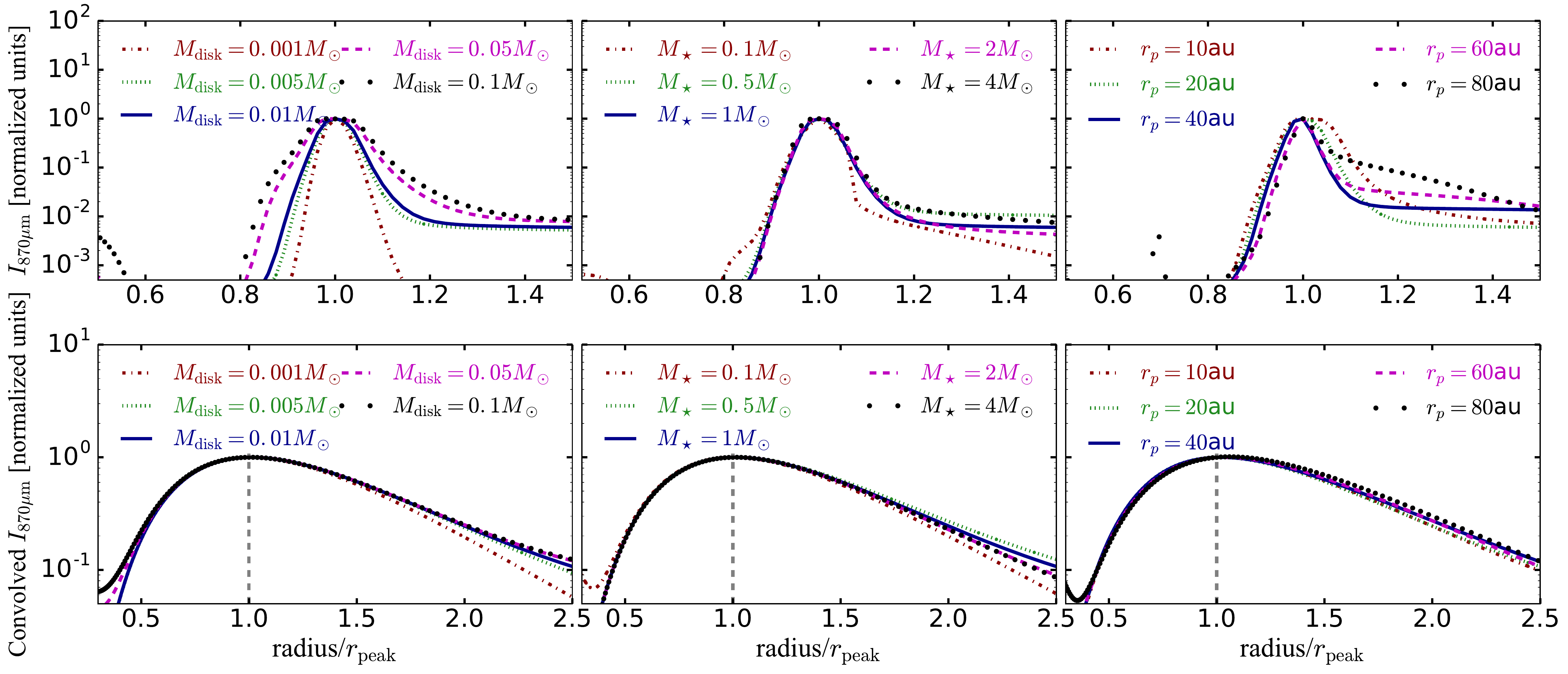}
	   \caption{Theoretical predictions of the intensity profiles using the dust density distributions of Fig.~\ref{model_predictions}. The difference between the upper and bottom panels is that the bottom are convolved with a 20\,au Gaussian beam profile, which is the typical resolution of the current observations. }
   \label{intensity_predictions}
\end{figure*}

Dust evolution models of particle trapping that invoke massive planets for the creation of pressure bumps can be used to predict correlations between the shape of the ring-like emission and the disk and stellar parameters. Figure~\ref{model_predictions} shows the dust density distribution of particles whose size is between 0.3 and 3\,mm (which is the range of grain-sized emitting at the wavelength of the observations we analyze in this work), in the case where one Jupiter mass planet is embedded in the disk. For these models, we assume a disk extension from 1 to 300\,au (logarithmically scaled radial grid) and the unperturbed gas surface density is given by $\Sigma_0 (r/r_p)^{-1}$, where $r_p$ is the planet position and $\Sigma_0$ controls the disk mass. The perturbation of the planet on the gas surface density (and hence on the pressure profile) is done assuming the analytical formulas in \cite{crida2006}, which accounts for the balance between the gravitational torque, pressure torque, and viscous torque to provide a gap shaped as a function of the disk and planet parameters. The depth of the gaps from this solution are corrected by the results in \cite{fung2014}, following the procedure described in \cite{pinilla2015c}. For the dust models, we assumed that all the grains are initially 1\,micron-sized and they are distributed as the gas density assuming a dust-to-gas ratio of 1/100. The dust density distribution evolves with time due to collisions and dynamics of the particles. We therefore include growth, fragmentation, and erosion of particles. For the dynamics of the dust grains, we take into account the drag with the gas (that provides the radial drift), and the turbulent diffusion as explained in \cite{binstiel2010}. For gas surface density, we assume an $\alpha_{\rm{turb}}-$ viscosity parameter of $10^{-3}$ \citep{shakura1973}, and hence the turbulent gas viscosity is $\nu=\alpha_{\rm{turb}}c_s^2 \Omega^{-1}$, where $c_s$ is the isothermal sound speed and $\Omega$ the Keplerian frequency. The dust diffusion is assumed to be as the turbulent gas viscosity \citep{youdin2007}, and the dust turbulent velocities are proportional to the square root of $\alpha_{\rm{turb}}$ as defined in \cite{ormel2007}. The standard model assumes a disk around one Solar mass, with a disk mass of 0.01\,$M_\odot$, and one Jupiter planet at 20\,au. We explored different values of the stellar mass, disk mass, and planet position, keeping all the rest of the parameters fixed, to explore how the ring-like shape accumulation of millimeter particles changes with these parameters.  All results are shown at the same time of evolution ($\sim$1\,Myr) in Fig.~\ref{model_predictions}. 

The top panel of Fig.~\ref{model_predictions} shows the results of the dust evolution models when the disk mass varies. The outer tail of the dust density distribution increases for higher disk mass. This is because for the same grain sizes (in this case $[0.3, 3]\,$mm), the coupling of the particles increases when the disk mass increases, and as a consequence because these particles are more coupled to the gas in higher disk masses, their radial drift velocities are lower \citep[e.g.][]{brauer2008}. The dust density distribution in the accumulation inside the pressure bump is similar in size for all cases, but it becomes narrower for the lowest mass disk that is considered, as expected because the radial drift for these grain sizes is the highest.

The middle panel of Fig.~\ref{model_predictions} shows the results when the stellar mass varies. As demonstrated in \cite{pinilla2013}, the drift of the particles increases in disks around low mass stars ($v_{\rm{drift}}\propto1/\sqrt{M_\star}$). This is reflected in the tail of the distribution of the millimeter-sized particles, which shrinks around low mass stars. 

Finally, the bottom panel of Fig.~\ref{model_predictions} shows the  dust density distribution when the planet is located at different radii. Locating the planet at different positions has two main effects. First, when the planet is located farther out, where the gas surface density is lower, it is expected that the coupling of the particles decreases and hence the radial drift of the particles increases, making the accumulation of dust in the pressure bump to narrow. However, as the planet is located farther out, the gap and the width of the pressure bump also increases; this can be reflected in a more wider accumulation of dust particles. As a consequence of these two effects, the potential correlation between the shape of dust accumulation and the position of the planet is not straightforward.  When the external width of the ring-like emission is normalized to the the peak, it is expected that it increases with the stellar and disk mass, suggesting a positive correlation between $\sigma_{\rm{ext}}/r_{\rm{peak}}$ and the stellar and disk mass, independent of the size of the cavity (or where the planet is located).

These results are also time dependent, since at longer times of evolution, the accumulation of dust particles also becomes narrower \citep{pinilla2015b}. But all depends on when the planet is formed in the disk \citep{pinilla2015c}  and on the planet mass \citep[e.g.][]{dejuanovelar2013}, which are a very hard properties to disentangle and constrain from observations of embedded planets \citep[see, e.g., the case of the potential planetary candidate in HD\,100546][]{quanz2013}.

To test if these correlations are observable at the millimeter emission, we calculate the intensity radial profile using the vertically integrated dust density distribution $\sigma(r,a)$ from the dust evolution models. At a given wavelength, the intensity is calculated as $I_\lambda(r)=B_\lambda (T(r)) [1 -\exp{(-\tau_\lambda (r))}]$, where $\tau_\lambda$ is the optical depth, which is computed as $\tau_\lambda=\sigma(r,a)\kappa_\lambda/\cos i$, where the opacities at a particular wavelength $\kappa_\lambda$ are calculated for each grain size as in \cite{pinilla2015c}-that is, assuming Mie theory and a mix of magnesium-iron silicates \citep{dorschner1995}. The temperature is assumed to be a simple power law that depends on the stellar properties as in \cite{kenyon1987}.

The intensity profiles calculated at $870\,\mu$m for each model are shown in Fig.~\ref{intensity_predictions}. We also convolved the intensity profile with a 20\,au Gaussian beam, which is the typical resolution of the observations presented in this work. The most clear correlation from the unconvolved profiles is with the disk mass, in which the ring-like emission becomes narrower for a lower disk mass. The other two correlations are very weak, and in the case of the planet position, there is not a clear relation between the planet position and the ring width, as discussed above. From the convolved intensity profiles (bottom panel of Fig.~\ref{intensity_predictions}), it is impossible to discern between the models. Higher angular resolution observations (with around 2-5\,au resolution) are needed to discern between these models, in which the external width of the ring-like emission increases with stellar and disk mass. The only clear result from the convolved profiles is that the ring of emission is radially asymmetric, with $\sigma_{\rm{ext}}>\sigma_{\rm{int}}$ in all cases. 

%%%
\subsection{Cavity size and ring shape correlations from observations} \label{sect:cav_correlations}
%%%

%%%%%%%%%%%%
%FIGURE 
%%%%%%%%%%%%
\begin{figure*}
 \centering
	\includegraphics[width=1.0\textwidth]{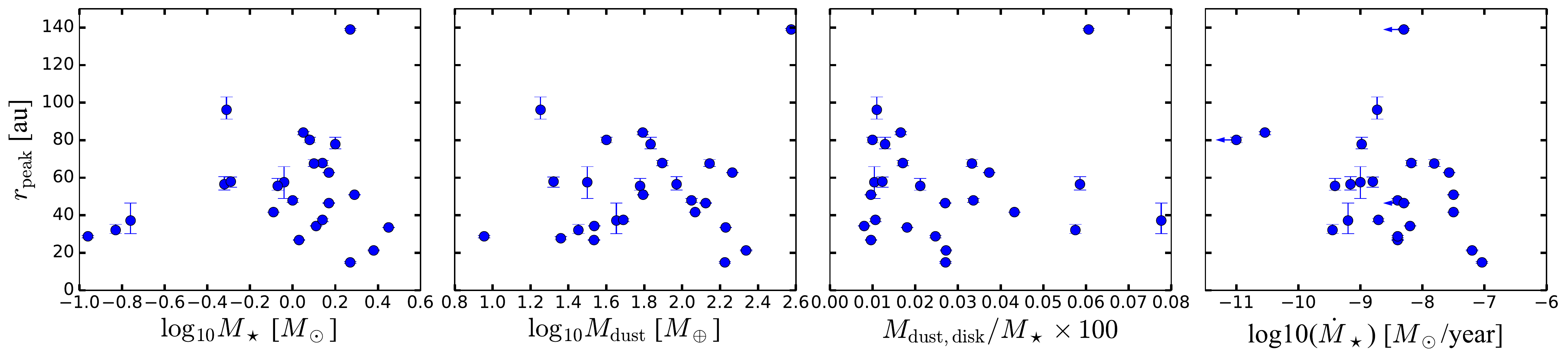}
	\caption{From left to right: correlations of the cavity size ($r_{\rm{peak}}$) with the stellar mass, disk dust mass, disk dust mass normalized with the stellar mass, and disk accretion rate for the TDs of our sample.}
   \label{peak_correlations}
\end{figure*}

We demonstrated in the previous sub-section that with the current resolution of the observations of most of the sources in our sample, we may not able to have any of the expected correlations from models of planet-disk interaction. Nevertheless, based on our MCMC results (Sect.~\ref{sect:analysis}), we look for any potential correlations of the cavity size ($r_{\rm{peak}}$) and the stellar and disk parameters. In particular, we are interested on any relation of $r_{\rm{peak}}$ with the stellar mass, disk (dust) mass, and disk accretion rate. With more future constraints from Gaia will also be interesting to look for correlations with the disk age. 

The cavity size ($r_{\rm{peak}}$) inferred in this work is similar as that previously reported for most of the disks \citep[e.g.][]{vandermarel2018}. For HD\,34282, \cite{vanderplas2017b} reported the inner edge of the cavity  at 80\,au, and the peak of the radial emission at 143\,au, in fair agreement with our results.  

One of the most clear results from models is that the ring of emission is radially asymmetric with a wider outer tail compared to the inner width of the ring. This is indeed the case for most of our targets (see Table~\ref{table:all_disks2}). Excluding RY\,Tau for which the cavity size is not well constrained, only three disks have $\sigma_{\rm{ext}}\ll\sigma_{\rm{int}}$ (i.e., a truncated outer disk that is not expected from the models of trapping by embedded planets). These three targets (Sz\,123A, J15583692, and CIDA1) do not share any particular property; they have different stellar/disk mass and accretion rates, and it is not clear why radial drift would be so efficient in these particular cases to create a truncated outer width for the ring of emission. A possibility is that an external companion or an encounter may had potentially truncated the outer disk. A few more disks (SR\,21 and $\rho$Oph\,3) have $\sigma_{\rm{ext}}/\sigma_{\rm{int}}\lesssim1$ (i.e., $\sim20\%$ of the disks analyzed in our sample do not follow the trends of dust evolution models and trapping with embedded planets previously described in Sect~\ref{sect:model_predic}).

The relation between $r_{\rm{peak}}$ with the stellar mass, disk (dust) mass (and also normalized by the mass of the central star), and disk accretion rate is shown in Fig.~\ref{peak_correlations}. This figure shows that from the current sample there is no clear correlation between the cavity size and these disk and stellar parameters. There are, however, some deserted areas. The most clear result is that there are not TDs with large cavities ($\gtrsim$45\,au) around low mass stars ($\lesssim$0.4\,$M_\odot$). However, for more massive stars, the cavity size can span one order of magnitude (from $\sim$10s to $\sim$100s  of au). This can be the consequence of smaller disks around very low mass stars \citep[e.g.][]{hendler2017a}, or that the sensitivity of current ALMA observations is not enough to detect large cavities around very low mass stars. 

In the case of a relation of $r_{\rm{peak}}$ with the disk dust mass, there is not a clear deserted area. Assuming a dust-to-gas ratio of 1/100, the disk mass of these TDs is for all cases lower than $\sim0.05\,M_\odot$, except for the case of HD\,34282, which is a clear outlier in this plot. 

When comparing  $r_{\rm{peak}}$ with the disk dust mass normalized to the central stellar mass,  we find similar results (i.e., no correlation with the cavity size). We also checked the correlation of $r_{\rm{peak}}-L_{\rm{mm}}$, with$L_{\rm{mm}}$ as the continuum luminosity, following observational results obtained in the Lupus star forming region \citep{tazzari2017a} and in a collection of disks of different regions by  \cite{tripathi2017}. In \cite{tripathi2017}, they found a strong correlation between the disk sizes (or $R_{\rm{eff}}$) and luminosities, such that $R_{\rm{eff}}\propto L_{\rm{mm}}^{0.5}$, and suggested that grain growth and the radial drift of particles can account for the observed trend. Alternatively,  optically thick emission can also explain the correlation. In our case, there is not a significant correlation between $r_{\rm{peak}}$ and $L_{\rm{mm}}$, or $r_{\rm{out}}$ and $L_{\rm{mm}}$ (defining $r_{\rm{out}}$ as $r_{\rm{peak}}+\rm{FWHM}_{\rm{ext}}$, with $\rm{FWHM}_{\rm{ext}}=2\sqrt{2\ln 2}\sigma_{\rm{ext}}$). The lack of a correlation between these quantities may also originate from the lack of radial drift inside a pressure trap, since as demonstrated by \cite{tripathi2017}, the trend can originate from the inward radial drift of pebbles. Thus, if radial drift is suppressed in pressure bumps, a correlation between $L_{\rm{mm}}$ and $r_{\rm{out}}$ (or $r_{\rm{peak}}$) is not expected. 

In the case of the potential relation of the cavity size with accretion rate, there is not clear correlation. Again, the TD around HD\,34282 is an outlier in this plot, but we only have an upper limit for the accretion rate. The span of the data in this plot differs from the relations expected from photoevaporation showed in, for example, \cite{owen2012, owen2017}, who reported that photoevaporation creates small cavities  
($\lesssim10\,$au) with low accretion rates ($\lesssim10^{-9}\,M_\odot$\,yr$^{-1}$).  As a result, we exclude photoevaporation as the leading mechanism of cavity formation for all the targets in our sample. This picture needs confirmation with higher angular resolution observations that can resolve smaller cavities,  more homogeneous measurements of accretion rates, and observations of the atomic content of carbon and oxygen in these disks \citep{ercolano2018}. The carbon abundance has been observed in only one of the targets of our sample, HD\,100546, for which the atomic carbon emission is as the interstellar gas-phase carbon abundance or depleted by very little \citep{kama2016}.

Dead zones and embedded planets remain as a possibility for the cavity formation for the disks in our sample, and for the lack of clear correlations between the cavity size and the stellar and disk parameters. It is important to note that the models of dust trapping by an embedded planet predict that this ring-like emission will become narrower with time. Knowing the disk distances with higher precision, and hence having a more constrained age for each disk, will allow us to test this idea in the future. 

Contrary to the models of dust trapping by embedded planets, in the models of particle trapping by dead zones, such an outer tail in the ring-like emission is not predicted \citep[see Fig.~8, from][]{pinilla2016}, and instead in this case the ring is much more radially symmetric at different times of evolution. Different disks in our sample show very radially symmetric rings (see Fig.~\ref{all_intensity_models}), as, for example J16042165. Therefore, the combination of future observations of TDs with ALMA at higher angular resolution, in combination with better constraints on disk ages, can help to discern between planet origin or dead zones.

In both cases, dead zones or planets, we expect that the  ring-like emission at longer wavelengths is narrower, because larger grains are more affected by radial drift and particle trapping. However, distinguishing between models at longer wavelengths requires higher resolution and longer observing time because of the fainter emission and the potentially narrower ring-like structure.

%%%%%%%%%%%%
%FIGURE 
%%%%%%%%%%%%
\begin{figure}
 \centering
	\includegraphics[width=1.0\columnwidth]{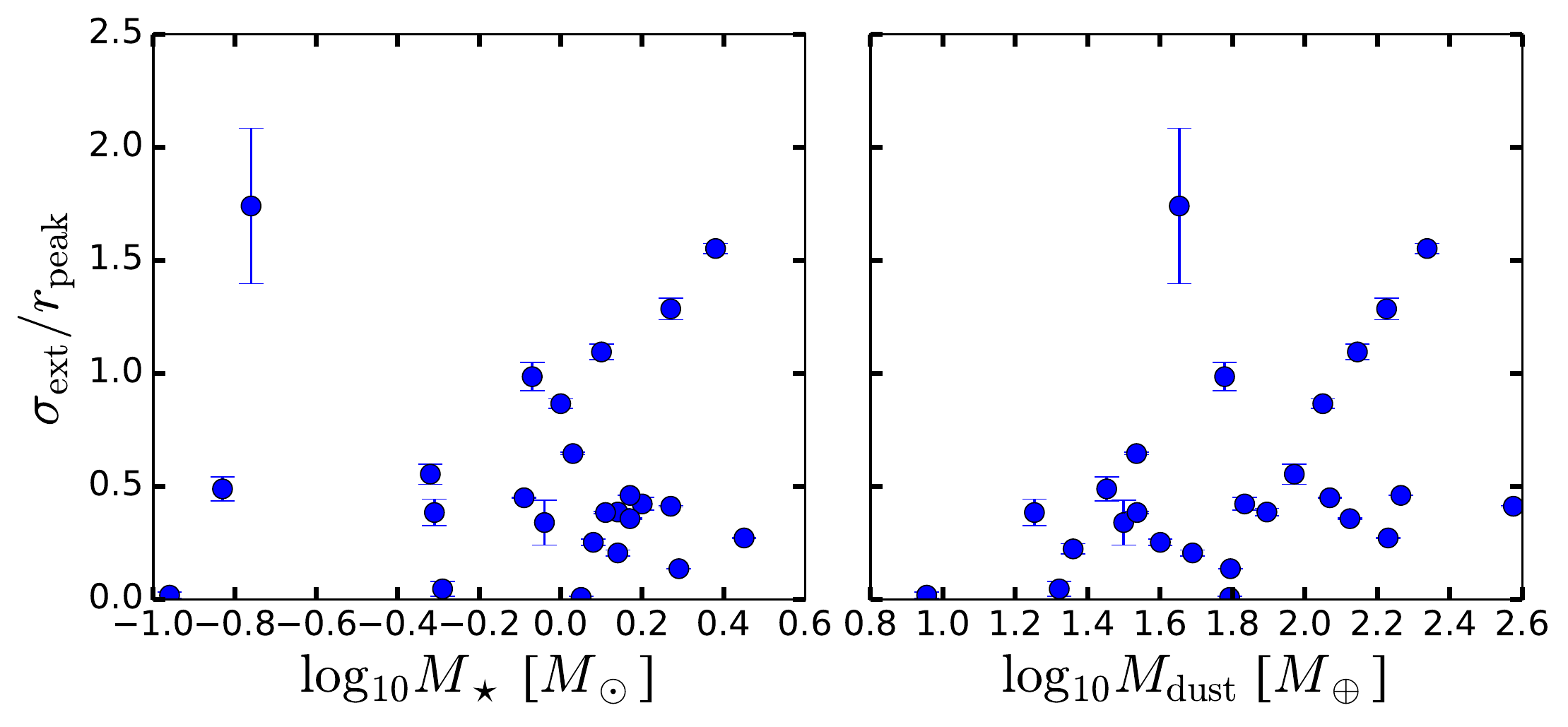}
	\caption{From left to right: correlations of the external width normalized to the peak of the cavity with the stellar mass and disk dust mass for the TDs of our sample.}
   \label{external_width_correlations}
\end{figure}

Even though the models predict that no correlation exists between the stellar/disk mass and the external width of the ring-like shape emission at the current resolution of the observations (Fig.~
\ref{intensity_predictions}), we check these correlations with the current data. Figure~\ref{external_width_correlations} shows the external width normalized to the peak location as a function of the stellar mass and dust disk mass.  These plots confirm our predictions from the models, without any clear correlation. However, in these relations, there is an almost deserted area where disks around low mass stars ($\lesssim0.8\,M_\odot$) and low disk (dust) mass ($\log_{10}M_{\rm{dust}}\lesssim1.8\,M_\oplus$) do not show wide rings with $\sigma_{\rm{ext}}/r_{\rm{peak}}>1$ (with the exception of J160708). These relations follow the theoretical prediction that drift is more efficient around low mass stars and in less massive disks (Fig.~\ref{model_predictions}), which narrows the ring-like emission.  

%%%%%%%%%%%%%%%%%%%%%%%%
\section{Conclusions} \label{sect:conclusions}
%%%%%%%%%%%%%%%%%%%%%%%%

In this paper, we  analyze the dust morphology of a total of 29 TDs that have been observed with ALMA, in order to characterize their cavity size and ring morphology. The sample excludes disks with multiple rings or gaps, or disks with strong azimuthal asymmetries, and we found three disks in our sample with yet unclassified millimeter-substructures ($\rho$Oph\,38, RXJ\,1615, and DM\,Tau) that are also excluded for the analysis. From this homogeneous analysis, we find a large diversity of cavity sizes and ring-like structures (Fig.~\ref{all_intensity_models}). We study these results in the context of different physical processes for the origin of the cavities in TDs. Our findings are summarized as follows.

\begin{enumerate}
\item The $M_{\rm{dust}}-M_\star$ relation is much flatter for TDs than the observed trends from samples of different star forming regions. We propose two potential reasons for a flatter relation. First, the emission is optically thick, which could preferentially affect the most massive disks. Second, particles are trapped in pressure maxima, which decreases or suppresses their radial drift. These two reasons are not necessarily exclusive. Models of dust evolution that include radial drift can explain the steepness of the $M_{\rm{dust}}-M_\star$ relation in different star forming regions \citep{pascucci2016}. However, in the case of TDs and trapping, the $M_{\rm{dust}}-M_\star$ is seen to be flatter. Based on our calculations of the optical depth at the peak of the ring-like emission, it is likely that the flatness of the $M_{\rm{dust}}-M_\star$ relation is a combination of the two reasons (optical thickness and trapping of particles). These possible explanations may also be the reason for a lack of a trend in the $r_{\rm{peak}}-L_{\rm{mm}}$ relation, which is steep in other samples \citep[e.g.][]{tazzari2017a, tripathi2017}. The  $M_{\rm{dust}}-M_\star$ relation also shows that cavities form in high (dust) disk mass, independent of the stellar mass.

\item Based on our results, we look for potential correlations between the cavity size of TDs and their stellar and disk properties. We find that there are not TDs with very large cavities ($\gtrsim$45\,au) around low mass stars ($\lesssim$0.4\,$M_\odot$). However, for more massive stars, the cavity size can span one order of magnitude (from $\sim$10s to $\sim$100s of au). This may be the consequence that disks are smaller around low mass stars. In addition, there is no trend between the cavity size and the dust disk mass. 

\item We also look for correlations between  the stellar/disk mass and the external width of the ring-like shape of emission. We found that disks around low mass stars ($\lesssim0.8\,M_\odot$) and low disk (dust) mass ($\log_{10}M_{\rm{dust}}\lesssim1.8\,M_\oplus$) do not show wide rings with $\sigma_{\rm{ext}}/r_{\rm{peak}}>1$. These relations follow the theoretical prediction that drift is more efficient around low mass stars and in less massive disks (Fig.~\ref{model_predictions}), which narrows the ring-like emission in the outer regions.  

\item We exclude photoevaporation as the leading mechanism for the cavity formation in the TDs of our sample, because there are not disks with small cavities and low accretion rates. However, the resolution of the observations in our sample is bias toward resolving large cavities. Higher angular resolution observations that can resolve smaller cavities, together with more homogeneous measurements of accretion rates, and observations of the atomic content of carbon and oxygen in these disks will discern if photoevaporation is still a possible explanation for a sub-set of TDs.

\item Models of giant embedded planets and dead zones remain as possible origin for the cavities in these TDs. One possible way to distinguish between these models is to better constrain the age of the individual disks. While dead zone models predict a radially symmetric ring-like structure at different times of evolution, models of embedded planets predict a radially asymmetric ring with a wider outer tail that becomes more symmetric with time. Therefore, if the inner and outer widths of the ring are similar in a young disk, the dead zone scenario is more likely than the planet scenario. Synergy between current/future observations at higher angular resolution observations of TDs with ALMA and better constraints of disks ages (with, e.g., Gaia), can help better constrain the models. In addition, observations that help constrain the total gas density distribution inside the millimeter cavities or that help constrain small (micron-sized) and intermediate (10s of microns) sized particles can provide a path to better understand the processes of disk dispersal that shape TDs. 

\end{enumerate}

 \software{CASA \citep{mcmullin2007}, GALARIO \citep{tazzari2018}, emcee \citep{foreman2013}}

\paragraph{Acknowledgments}
  \acknowledgments{Authors are very thankful to A.~Natta for all the discussions about the results of this paper, and with the referee for the constructive report. P.P. acknowledges support by NASA through Hubble Fellowship grant HST-HF2-51380.001-A awarded by the Space Telescope Science Institute, which is operated by the Association of Universities for Research in Astronomy, Inc., for NASA, under contract NAS 5-26555. M.T. has been supported by the DISCSIM project, grant agreement 341137 funded by the European Research Council under ERC-2013-ADG. M.T., C.F.M. and L.T. acknowledge support by the Deutsche Forschungsgemeinschaft (DFG, German Research Foundation) - Ref no. FOR 2634/1. I.P. and N.H. acknowledge support from an NSF Astronomy \& Astrophysics Research Grant (ID: 1515392). C.F.M. acknowledges support through an ESO Fellowship. G.v.d.P. acknowledges funding from ANR of France under contract number ANR-16-CE31-0013. S.A.B. acknowledges support from the National Science Foundation Graduate Research Fellowship under grant No. DGE1144469 and from NSF grant No. AST-1140063. In addition,  an allocation of computer time from the UA Research Computing High Performance Computing (HPC) at the University of Arizona to perform the simulations presented in this paper is gratefully acknowledged. This paper makes use of the following ALMA data: ADS/JAO.ALMA \#2011.0.00724.S, 2011.1.00863.S, 2011.0.00966.S, 2012.1.00158.S, 2012.1.00182.S, 2013.1.00091.S, 2013.1.00157.S, 2013.1.00220.S, 2013.1.00395.S, 2013.1.00437.S, 2013.1.00498.S, 2013.1.00658.S, 2013.1.00663.S, 2013.1.01020S,  and 2015.1.00934.S. ALMA is a partnership of ESO (representing its member states), NSF (USA) and NINS (Japan), together with NRC (Canada), MOST and ASIAA (Taiwan), and KASI (Republic of Korea), in cooperation with the Republic of Chile. The Joint ALMA Observatory is operated by ESO, AUI/NRAO and NAOJ. This work has made use of data from the European Space Agency (ESA) mission {\it Gaia} (\url{https://www.cosmos.esa.int/gaia}), processed by the {\it Gaia} Data Processing and Analysis Consortium (DPAC, \url{https://www.cosmos.esa.int/web/gaia/dpac/consortium}). Funding for the DPAC has been provided by national institutions, in particular the institutions participating in the {\it Gaia} Multilateral Agreement.}

\end{document}